\documentclass[final,12pt]{elsarticle}

\usepackage[margin=1in]{geometry}

\usepackage{graphicx}

\usepackage{amssymb}

\usepackage{url}

\usepackage{amsmath}
\usepackage{upgreek}

\usepackage{amsthm}

\usepackage{tensor}
\usepackage{dsfont}

\usepackage{siunitx}

\usepackage{amssymb}

\usepackage{bbm}

\usepackage{soul}

\usepackage{algorithmicx}
\usepackage{algorithm}
\usepackage[noend]{algpseudocode}
\makeatletter
\renewcommand{\ALG@beginalgorithmic}{\footnotesize}
\makeatother

\usepackage[dvipsnames,table,xcdraw]{xcolor}

\usepackage{booktabs}

\usepackage{amsmath}

\makeatletter
\newcommand{\vast}{\bBigg@{4}}
\newcommand{\Vast}{\bBigg@{7}}
\makeatother

\usepackage{caption}

\usepackage{subcaption}

\usepackage{bm}
\usepackage{amsbsy}
\usepackage{amsfonts}
\usepackage{stackrel}
\usepackage{mathrsfs} 
\usepackage{mathtools}

\usepackage{todonotes}
\usepackage{booktabs}
\usepackage{adjustbox}

\newcommand{\pp}[2]{ \frac{\partial #1}{\partial #2} }

\newcommand{\DDt}[1]{ \frac{D #1}{Dt} }

\setcitestyle{authoryear}
 \biboptions{comma,round}

\journal{Journal of Computational Physics}
\usepackage{etoolbox}

\begin{document}

\begin{frontmatter}

\title{Assessment of diffuse-interface methods for compressible\\ multiphase fluid flows and elastic-plastic deformation in solids}



\author{\underline{Suhas S. Jain} \footnote[1]{equal contribution}}
\ead{sjsuresh@stanford.edu}
\author{\underline{Michael C. Adler} \footnotemark}
\ead{mcadler@stanford.edu}

\author{\underline{Jacob R. West} \footnotemark[1]}
\ead{jrwest@stanford.edu}

\author{Ali Mani}
\author{Parviz Moin}
\author{Sanjiva K. Lele}
\address{Center for Turbulence Research, Stanford University, California, USA 94305}


\begin{abstract}

This work describes three diffuse-interface methods for the simulation of immiscible, compressible multiphase fluid flows and elastic-plastic deformation in solids.
The first method is the \textit{localized-artificial-diffusivity approach} of \citet{cook2007artificial}, \citet{subramaniam2018high}, and \citet{adler2019strain}, in which artificial diffusion terms are added to the individual phase mass fraction transport equations and are coupled with the other conservation equations.
The second method is the \textit{gradient-form approach} that is based on the quasi-conservative method of \citet{shukla2010interface}, in which the diffusion and sharpening terms (together called regularization terms) are added to the individual phase volume fraction transport equations and are coupled with the other conservation equations \citep{tiwari2013diffuse}.
The third approach is the \textit{divergence-form approach} that is based on the fully conservative method of \citet{jain2020conservative}, in which the regularization terms are added to the individual phase volume fraction transport equations and are coupled with the other conservation equations. 
In the present study, all three diffuse-interface methods are used in conjunction with a four-equation, multicomponent mixture model, in which pressure and temperature equilibria are assumed among the various phases.

The primary objective of this work is to compare these three methods in terms of their ability to: maintain constant interface thickness throughout the simulation; conserve mass, momentum, and energy; and maintain accurate interface shape for long-time integration. The second objective of this work is to consistently extend these methods to model interfaces between solid materials with strength. To assess and compare the methods, they are used to simulate a wide variety of problems, including (1) advection of an air bubble in water, (2) shock interaction with a helium bubble in air, (3) shock interaction and the collapse of an air bubble in water, and (4) Richtmyer--Meshkov instability of a copper--aluminum interface. The current work focuses on comparing these methods in the limit of relatively coarse grid resolution, which illustrates the true performance of these methods. This is because it is rarely practical to use hundreds of grid points to resolve a single bubble or drop in large-scale simulations of engineering interest.

\end{abstract}


\end{frontmatter}



\section{Introduction}


Compressible multiphase fluid flow and multiphase elastic-plastic deformation of solid materials with strength are important phenomena in many engineering applications, including shock compression of condensed matter, detonations and shock-material-interface interactions, impact welding, high-speed fuel atomization and combustion, and cavitation and bubble collapse motivated by both mechanical and biomedical systems. In this work, we are concerned with the numerical modeling of multiphase systems, i.e., those systems that involve two or more phases of gas, liquid, or solid in the domain. The numerical simulation of these multiphase systems presents several new challenges in addition to those associated with analogous single-phase simulations.
These modeling complications include but are not limited to (1) representing the phase interface on an Eulerian grid; (2) resolving discontinuities in quantities at the interface, especially for high-density ratios; (3) maintaining conservation of (a) the mass of each phase, (b) the mixture momentum, and (c) the total energy of the system; and (4) achieving an accurate mixture representation of the interface for maintaining thermodynamic equilibria.
Hence, the numerical modeling of multiphase compressible fluid flows and deformation of solid materials with strength are still an active area of research.

With these numerical challenges in mind, we choose to pursue the single-fluid approach \citep{kataoka1986local}, in which a single set of equations is solved to describe all of the phases in the domain, as opposed to a multi-fluid approach, which requires solving a separate set of equations for each of the phases.
We are presented with various choices in terms of the system of equations that can be used to represent a compressible multiphase system.
In this work, we employ a multicomponent system of equations (a four-equation model) that assumes spatially local pressure and temperature equilibria, including at locations within the diffuse material interface \citep{shyue1998efficient, venkateswaran2002computation,marquina2003flux,cook2009enthalpy}.
Relaxing the assumption of temperature equilibrium, \citet{allaire2002five} and \citet{kapila2001two}
developed the five-equation model that has proven successful for a variety of applications with high density ratios, strong compressibility effects, and phases with disparate equations of state (EOS), and has been widely adopted for the simulation of compressible
two-phase flows
\citep{shukla2010interface, so2012anti, ansari2013numerical, shukla2014nonlinear, coralic2014finite, tiwari2013diffuse, perigaud2005compressible, wong2017high, chiapolino2017sharpening, garrick2017interface, garrick2017finite, Jain2018, jain2020conservative}.
%
Finally, there are six- and seven-equation models that are more general and include more non-equilibrium effects but are not as widely used for the simulation of two-phase flows \citep{yeom2013modified,baer1986two,sainsaulieu1995finite,saurel1999simple}.

For representing the interface on an Eulerian grid, we use an interface-capturing method, as opposed to an interface-tracking method, due to the natural ability of the former method to simulate dynamic creation of interfaces and topological changes \citep{mirjalili2017interface}. Interface-capturing methods can be classified into sharp-interface and diffuse-interface methods.
In this work, we choose to use diffuse-interface methods for modeling the interface between compressible materials  \citep{saurel2018diffuse}.
This choice is due to the natural advantages that the diffuse-interface methods offer over the sharp-interface methods, such as ease of representation of the interface, low cost, good conservation properties, and parallel scalability.


Historically, diffuse-interface methods for compressible flows involved modeling the interface implicity, i.e., with no explicit interface capturing through regularization/reconstruction. These methods can be classified as implicit diffuse-interface methods. These methods assume that the underlying numerical methods are capable of handling the material interfaces, a concept similar to implicit large-eddy simulation. One challenge with the implicit diffuse-interface capturing of material interfaces is that the interface tends to diffuse over time. Unlike shock waves, in which the convective characteristics sharpen the shock over time, material interfaces (like contact discontinuities) do not sharpen naturally; therefore, modeling material interfaces requires an active balance between interface sharpening and diffusion to maintain an appropriate interface thickness over time. Therefore, in the present work, the focus is on explicit diffuse-interface methods. These methods explicitly model the interface using the interface regularization/reconstruction techniques.


This paper explores three explicit
diffuse-interface methods that are representative of the different approaches for this problem and possesses unique characteristics.
The first approach (referred to as the LAD approach) is based on the localized-artificial-diffusivity (LAD) method~\citep{cook2007artificial,subramaniam2018high,adler2019strain}, in which localized, nonlinear diffusion terms are added to the individual phase mass transport equations and coupled with the other conservation equations.
This method conserves the mass of individual phases, mixture momentum, and total energy of the system due to the conservative nature of the diffusion terms added to the system of equations and results in no net mixture-mass transport. 
This method is primarily motivated by applications involving miscible, multicomponent, single-phase flows, but it has been successfully adapted for multiphase flow applications.
The idea behind this approach is to effectively add species diffusion in the selected regions of the domain to properly resolve the interface on the grid and to prevent oscillations due to discontinuities in the phase mass equations.
High-order compact derivative schemes can be used to discretize the added diffusion terms without resulting in distortion of the shape of the interface over long-duration time advancement.
However, one drawback of this approach is that the interface thickness increases with time due to the lack of sharpening fluxes that act against the diffusion.
This method is therefore most effective for problems in which the interface is in compression (such as shock/material-interface interactions with normal alignment).
However, the deficiency of this method due to the lack of a sharpening term is evident for applications in which the interface between immiscible materials undergoes shear or expansion/tension.
LAD formulations have also been examined in the context of five-equation models, in which localized diffusion is also added to the volume fraction transport equation \citep{aslani2018localized}.

The second approach (referred to as the gradient-form approach) is based on the quasi-conservative method proposed by \citet{shukla2010interface}, in which diffusion and sharpening terms (together called regularization terms) are added for the individual phase volume fraction transport equations and coupled with the other conservation equations \citep{tiwari2013diffuse}.
This method only approximately conserves the mass of individual phases, mixture momentum, and total energy of the system due to the non-conservative nature of the regularization terms added to the system of equations.
In contrast to the LAD approach, this method can result in net mixture-mass transport, which can sharpen or diffuse the mixture density; depending on the application, this may be an advantageous or disadvantageous property.
The primary advantage of this method is that the regularization terms are insensitive to the method of discretization; they can be discretized using high-order compact derivative schemes without distorting the shape of the interface over long-duration time advancement.
However, the non-conservative nature of this approach results in poor performance of the method for certain applications.
For example, premature topological changes and unphysical interface behavior can be observed when the interfaces are poorly resolved (exacerbating the conservation error) and subjected to shocks that are not aligned with the interface. 

The third approach (referred to as the divergence-form approach) is based on the fully conservative method proposed by \citet{jain2020conservative}, in which diffusion and sharpening terms are added to the individual phase volume fraction transport equations and coupled with the other conservation equations.
This method conserves the mass of individual phases, mixture momentum, and total energy of the system due to the conservative nature of the regularization terms added to the system of equations.
Similar to the gradient-form approach and in contrast to the LAD approach, this method can result in net mixture-mass transport, which can sharpen or diffuse the mixture density.
The primary challenge of this method is that one needs to be careful with the choice of discretization used for the regularization terms.
Using a second-order finite-volume scheme (in which the nonlinear fluxes are formed on the faces), \citet{jain2020conservative} showed that a discrete balance between the diffusion and sharpening terms is achieved, thereby eliminating the spurious behavior that was discussed by \citet{shukla2010interface}.
The idea behind this is similar to the use of the balanced-force algorithm \citep{francois2006balanced,mencinger2007finite} for the implementation of the surface-tension forces, in which a discrete balance between the pressure and surface-tension forces is necessary to eliminate the spurious velocity around the interface.
The current study also demonstrates that appropriately crafted higher-order schemes may be used to effectively discretize the regularization terms.   
This method is free of premature topological changes and unphysical interface behavior present with the previous approach.
However, due to the method of discretization, the anisotropy of the derivative scheme can more significantly distort the shape of the material interface over long-duration time advancement in comparison to the gradient-form approach; the severity of this problem is significantly reduced when using higher-order schemes.

For all the three diffuse-interface methods considered in this work, it is important to include physically consistent corrections, associated with the interface regularization process, in each of the governing equations.
For example, \citet{cook2009enthalpy}, \citet{tiwari2013diffuse}, and \citet{jain2020conservative} discuss physically consistent regularization terms for the LAD, gradient-form, and divergence-form approaches, respectively.
The physically consistent regularization terms of \citet{cook2009enthalpy}, \citet{tiwari2013diffuse}, and \citet{jain2020conservative} are derived in such a way that the regularization terms do not spuriously contribute to the kinetic energy and entropy of the system.
This significantly improves the stability of the simulation, especially for flows with high density ratios.
However, discrete conservation of kinetic energy and entropy is needed to show the stability of the methods for high-Reynolds-number turbulent flows \citep{jain2020keep}. 

We employ a fully Eulerian method for modeling the deformation of solid materials, as opposed to a fully Lagrangian approach \citep{benson1992computational} or a mixed approach such as arbitrary-Lagrangian-Eulerian methods \citep{donea2004arbitrary}, because of its cost-effectiveness and accuracy to handle large deformations.
There are various Eulerian approaches in the literature that differ in the way the deformation of the material is tracked.
The popular methods employ the inverse deformation gradient tensor \citep{miller2001high, ortega2014numerical, ghaisas2018unified}, the left Cauchy-Green tensor \citep{sugiyama2010full,sugiyama2011full}, the co-basis vectors \citep{favrie2011mathematical}, the initial material location \citep{valkov2015eulerian,jain2019conservative}, or other variants of these methods to track the deformation of the material in the simulation. 
In this work we use the inverse deformation gradient tensor approach because of its applicability to model plasticity.
We propose consistent corrections in the kinematic equations, that describe the deformation of the solid, associated with the interface regularization process.


In summary, the two main objectives of this paper are as follows.
The first objective is to assess several diffuse-interface-capturing methods for compressible two-phase flows.
The interface-capturing methods in this work will be used with a four-equation multicomponent model; however, they are readily compatible with a variety of other models, including the common five-, six-, or seven-equation models.
The second objective is to extend these interface-capturing methods for the simulation of elastic-plastic deformation in solid materials with strength, including comparison of these methods in the context of modeling interfaces between solid materials.

The remainder of this paper is outlined as follows: Section \ref{sec:theory} describes the three diffuse-interface methods considered in this study, along with the details of their implementation. Section \ref{sec:results} discusses the application of these methods to a variety of problems including a shock/helium-bubble interaction in air, an advecting air bubble in water, a shock/air-bubble interaction in water, and a Richtmyer--Meshkov instability of an interface between copper and aluminum. Concluding remarks are made in Section~\ref{sec:conclusions} along with a summary. A table highlighting the strengths and limitations of the different methods considered in this work is also presented in this section.



\section{Theoretical and numerical model}\label{sec:theory}

\subsection{Governing equations}
The governing equations for the evolution of the multiphase flow or multimaterial continuum in conservative Eulerian form are described in Eqs. \eqref{eqn:mass}-\eqref{eqn:energy}. This consists of the conservation of species mass (Eq.~\ref{eqn:mass}), total momentum (Eq.~\ref{eqn:momentum}), and total energy (Eq.~\ref{eqn:energy}).
These are followed by the kinematic equations that track the material deformation, which include transport equations for the elastic component of the inverse deformation gradient tensor (Eq.~\ref{eqn:elastic_g}), and the plastic Finger tensor (Eq.~\ref{eqn:plastic_g}).
\begin{equation}\label{eqn:mass}
    \underbrace{\frac{\partial \rho Y_m}{\partial t}}_{\substack{\textrm{local} \\ \textrm{derivative}}} + \underbrace{\frac{\partial u_k \rho Y_m}{\partial x_k}}_{\textrm{advection}} = - \underbrace{\frac{\partial \left( J^*_m \right)_i}{\partial x_i}}_{\substack{\textrm{artificial} \\ \textrm{diffusion}}} + \underbrace{J_m}_{\substack{\textrm{interface} \\ \textrm{regularization}}},
\end{equation}
\begin{equation}\label{eqn:momentum}
    \underbrace{\frac{\partial \rho u_i}{\partial t}}_{\substack{\textrm{local} \\ \textrm{derivative}}} + \frac{\partial}{\partial x_k} \left( \underbrace{u_k \rho u_i}_{\textrm{advection}} - \underbrace{\sigma_{ik}}_{\substack{\textrm{stress} \\ \textrm{source}}} ~\right) = \underbrace{\frac{\partial \tau^*_{ik}}{\partial x_k}}_{\substack{\textrm{artificial} \\ \textrm{diffusion}}} + \underbrace{F_i}_{\substack{\textrm{interface} \\ \textrm{regularization}}},
\end{equation}
\begin{equation}\label{eqn:energy}
    \begin{aligned}
        \underbrace{\frac{\partial}{\partial t} \left[ \rho \left( e+ \frac{1}{2} u_j u_j \right) \right]}_{\substack{\textrm{local} \\ \textrm{derivative}}} &+ \frac{\partial}{\partial x_k} \left[ \underbrace{u_k \rho \left( e + \frac{1}{2} u_j u_j \right)}_{\textrm{advection}} - \underbrace{u_i \sigma_{ik}}_{\substack{\textrm{stress} \\ \textrm{source}}} \right] \\
        &= \underbrace{\frac{\partial}{\partial x_k} \left( u_i \tau^*_{ik} - q^*_k \right)}_{\textrm{artificial diffusion}} + \underbrace{H}_{\substack{\textrm{interface} \\ \textrm{regularization}}},
    \end{aligned}
\end{equation}
\begin{equation}\label{eqn:elastic_g}
    \begin{aligned}
    \underbrace{\frac{\partial g^e_{ij}}{\partial t}}_{\substack{\textrm{local} \\ \textrm{derivative}}} &+ \underbrace{\frac{\partial g^e_{ik} u_k}{\partial x_j}}_{\substack{\textrm{curl-free} \\ \textrm{advection/strain}}} + \underbrace{u_k \left( \frac{\partial g^e_{ij}}{\partial x_k} - \frac{\partial g^e_{ik}}{\partial x_j} \right)}_{\substack{\textrm{non-zero curl} \\ \textrm{advection/strain}}} - \underbrace{\frac{1}{2\mu\tau_{rel}} g^e_{ik} \sigma'_{kj}}_{\textrm{elastic-plastic source}} \\ &= \underbrace{\frac{\zeta^e}{\Delta t} \left( \frac{\rho}{\rho_0 \det{\left(\bm{g^e}\right)}}-1\right) g^e_{ij}}_{\textrm{density compatibility}} + \underbrace{\frac{\partial}{\partial x_k}\left( {g^e}^* \frac{\partial g^e_{ij}}{\partial x_k}\right)}_{\textrm{artificial diffusion}} + \underbrace{K^e_{ij}}_{\substack{\textrm{interface} \\ \textrm{regularization}}},
    \end{aligned}
\end{equation}

\begin{equation}\label{eqn:plastic_g}
    \begin{aligned}
    \underbrace{\frac{\partial G^p_{ij}}{\partial t}}_{\substack{\textrm{local} \\ \textrm{derivative}}} &+ \underbrace{u_k \frac{\partial G^p_{ij}}{\partial x_k}}_{\textrm{advection}} + \underbrace{\frac{1}{2\mu\tau_{rel}} \left( G^p_{ik} g^e_{kl} \sigma'_{lm} \left(g^e\right)^{-1}_{mj} +  G^p_{jk} g^e_{kl} \sigma'_{lm} \left(g^e\right)^{-1}_{mi} \right)}_{\textrm{elasto-plastic source}} \\ &= \underbrace{\frac{\zeta^p}{\Delta t} \left( \frac{1}{ \det{\left(\bm{G^p}\right)}^{1/2}}-1\right) G^p_{ij}}_{\textrm{density compatibility}} + \underbrace{\frac{\partial}{\partial x_k}\left( {g^p}^* \frac{\partial G^p_{ij}}{\partial x_k}\right)}_{\textrm{artificial diffusion}}.  
    \end{aligned}
\end{equation}
Here, $t$ and $\bm{x}$ represent time and the Eulerian position vector, respectively.
$Y_m$ describes the mass fraction of each constituent material, $m$. The variables $\bm{u}$, $\rho$, $e$, and $\bm{\underline{\sigma}}$ describe the mixture velocity, density, internal energy, and Cauchy stress, respectively, which are related to the species-specific components by the relations $\rho=\sum_{m=1}^M\phi_m\rho_m$, $e=\sum_{m=1}^MY_m e_m$, and $\bm{\underline{\sigma}}=\sum_{m=1}^M\phi_m\bm{\underline{\sigma}}_m$, in which $\phi_m$ is the volume fraction of material $m$, and $M$ is the total number of material constituents. The variables $g^e_{ij}$ and $G^p_{ij}$ are tensors that track elastic and plastic material deformation in problems with solids. These equations are described in greater detail in the next section. 

The right-hand-side terms describe the localized artificial diffusion (see also Section~\ref{sec:LAD}), including the artificial viscous stress, $\tau^*_{ik}=2\mu^*S_{ik} + \left(\beta^*-2\mu^*/3\right) \left(\partial u_j/ \partial x_j\right)\delta_{ik}$, and the artificial enthalpy flux, $q^*_i=-\kappa^*\partial T/\partial x_i + \sum_{m=1}^M h_m\left(J^*_m\right)_i$, with strain rate tensor, $S_{ik}=\left(\partial u_i/\partial x_k + \partial u_k/\partial x_i \right)/2$, and temperature, $T$.
The second term in the artificial enthalpy flux expression is the enthalpy diffusion term \citep{cook2009enthalpy}, in which $h_m=e_m+p_m/\rho_m$ is the enthalpy of species $m$.
The artificial Fickian diffusion of species $m$ is described by $\left(J^*_m\right)_i=-\rho\left[ D^*_m \left(\partial Y_m/\partial x_i \right) - Y_m \sum_k D^*_k \left(\partial Y_k/\partial x_i \right)\right]$.

\subsection{Material deformation and plasticity model}
The kinematic equations that describe the deformation of the solid in the Eulerian framework employ the inverse deformation gradient tensor, $g_{ij}=\partial X_i/\partial x_j$, in which $\bm{X}$ and $\bm{x}$ describe the position of a continuum parcel in the material (Lagrangian) and spatial (Eulerian) perspectives, respectively.
In this work, a single inverse deformation gradient is used to describe the kinematics of the mixture \citep{ghaisas2017evaluation, ghaisas2018unified}. 
Following \cite{miller2001high}, a multiplicative decomposition of the total inverse deformation gradient tensor, $\bm{\underline{g}}$, into elastic, $\bm{\underline{g^e}}$, and plastic, $\bm{\underline{g^p}}$, components is assumed, $g_{ij}=g^p_{ik}g^e_{kj}$, reflecting the assumption that the plastic deformation is recovered when the elastic deformation is reversed, $g^p_{ij}=g_{ik}\left(g^e\right)^{-1}_{kj}$.
It is additionally assumed that the plastic deformation is volume preserving \citep{plohr1992conservative}, providing compatibility conditions for the inverse deformation gradient tensor determinants, $\det{\left(\bm{\underline{g^p}}\right)}=1$ and $\det{\left(\bm{\underline{g}}\right)}=\det{\left(\bm{\underline{g^e}}\right)}=\rho/\rho_0$, in which $\rho_0$ represents the undeformed density and $\det{(\cdot)}$ represents the determinant operator. 
In this work, the plastic Finger tensor $G^p_{ij} = g^p_{ik} g^p_{jk}$ is solved for because it tends to be more stable than the equation for $\textbf{g}^p$, and because models for strain hardening are often parametrized in terms of norms of the plastic Finger tensor. This choice and its alternatives are discussed in detail in \cite{Adler_West_Lele_2021Manuscript_AHighOrder}. 

We also assume that the materials with strength are elastic perfectly plastic, i.e., the material yield stress is independent of strain and strain rate; thus, only the elastic component of the inverse deformation gradient tensor is necessary to close the governing equations. As a result, we solve only the equation for elastic deformation in the present work.
The plastic component of the inverse deformation gradient tensor, or the full tensor, can be employed to supply the plastic strain and strain rate necessary for more general plasticity models~\citep{adler2019strain}. 

Plastic deformation is incorporated into the numerical framework by means of a visco-elastic Maxwell relaxation model, which has been employed recently in several Eulerian approaches \citep{ndanou2015multi, ortega2015richtmyer, ghaisas2018unified}.
The plastic relaxation timescale is described by
\begin{equation}
    \frac{1}{\tau_{\textrm{rel}}} = \frac{1}{\left(\rho/\rho_0\right)\tau_0}\left[ \frac{\textrm{R}\left(\left\lVert \bm{\underline{\sigma}}'\right\rVert^2 -\frac{2}{3}\sigma_Y^2\right)}{\mu^2}\right],
\end{equation}
in which $\bm{\underline{\sigma}}'=\textrm{dev}\left(\bm{\underline{\sigma}}\right)$ and $\mu$ is the material shear modulus.
The ramp function $R\left(x\right)=\textrm{max}\left(x,0\right)$ turns on plasticity effects only when the yield criterion is satisfied.
In many cases, the elastic-plastic source term is stiff due to the small value of $\tau_{\textrm{rel}}$ relative to the convective deformation scales.
To overcome this time step restriction, implicit plastic relaxation is performed at each timestep, based on the method of \cite{favrie2011mathematical} and described by \cite{ghaisas2018unified}.

\subsection{Equations of state and constitutive equations}
A hyperelastic constitutive model, in which the elastic stress--strain relationship is compatible with a strain energy-density functional, is assumed to close the thermodynamic relationships in the governing equations.
The internal energy, $e$, is additively decomposed into a hydrodynamic component, $e_h$, and an elastic component, $e_e$, as in \cite{ndanou2015multi}.
The hydrodynamic component is analogous to a stiffened gas, with
\begin{equation}
    e = e_h\left(p,\rho\right) + e_e\left(\bm{\underline{\hat{g}}}\right), \qquad e_h = \frac{p+\gamma p_{\infty}}{\left(\gamma-1 \right)\rho}, \qquad e_e = \frac{\mu}{4 \rho_0} \textrm{tr}\left[\left( \bm{\underline{\hat{g}}} - \bm{\underline{I}} \right)^2\right],
\end{equation}
in which $\bm{\underline{\hat{g}}} = \det{\left( \bm{\underline{G}^e} \right)}^{-1/3} \bm{\underline{G}^e}$, $\bm{\underline{G}^e} = \bm{\underline{g}^e{}}^T\bm{\underline{g}^e}$, $p$ is the pressure, $p_{\infty}$ (with units of pressure) and $\gamma$ (nondimensional) are material constants of the stiffened gas model for the hydrodynamic component of internal energy. 
With this EOS, the Cauchy stress, $\bm{\underline{\sigma}}$, satisfying the Clausius-Duhem inequality is described by
\begin{equation}
    \bm{\underline{\sigma}} = -p \bm{\underline{I}} - \mu \frac{\rho}{\rho_0} \left\{ \det{ \left( \bm{\underline{G}^e}\right) ^{-2/3}} \textrm{dev} \left[ \left( \bm{\underline{G}^e} \right)^2 \right] - \det{\left(\bm{\underline{G}^e}\right)}^{-1/3} \textrm{dev} \left( \bm{\underline{G}^e} \right) \right\},
\end{equation}
in which $\textrm{dev} \left( \bm{\underline{G}^e} \right)$ signifies the deviatoric component of the tensor: $\textrm{dev} \left( \bm{\underline{G}^e} \right) = \bm{\underline{G}^e} - \frac{1}{3}\textrm{tr} \left( \bm{\underline{G}^e} \right) \bm{\underline{1}}$, with $\textrm{tr} \left( \cdot \right)$ signifying the trace of the tensor and $\bm{\underline{1}}$ signifying the identity tensor.
The elastic component of the internal energy, $\epsilon_e$, is assumed to be isentropic.
Therefore, the temperature, $T$, and entropy, $\eta$, are defined by the hydrodynamic stiffened gas component of the EOS, as follows.
\begin{equation}
    \begin{aligned}
    &e_h = C_v T \left( \frac{p+\gamma p_{\infty}}{p+p_{\infty}}\right), \qquad R = C_p -C_v, \qquad \gamma = \frac{C_p}{C_v}, \\
    &\eta - \eta_0 = C_p \ln{\left( \frac{T}{T_0}\right)} + R \ln{\left( \frac{p_0 + p_{\infty}}{p+p_{\infty}} \right)}.
    \end{aligned}
\end{equation}
Here, $\eta_0$ is the reference entropy at pressure, $p_0$, and temperature, $T_0$. In the case of compressible flow with no material strength, the material model reduces to the stiffened gas EOS commonly employed for liquid/gas-interface interactions~\citep{shukla2010interface,jain2020conservative}.

\subsection{Pressure and temperature equilibration method \label{sec:equilibrium}}
Many models for multiphase simulation assume that the thermodynamic variables are not in equilibrium, necessitating the solution of an additional equation for volume fraction transport~\citep{shukla2010interface,jain2020conservative}.
Our model begins with the assumption that both pressure and temperature remain in equilibrium between the phases.
The equilibration method follows from \citet{cook2009enthalpy} and \citet{subramaniam2018high}.
For a mixture of $M$ species, we solve for $2M+2$ unknowns, including the equilibrium pressure $\left( p \right)$, the equilibrium temperature $\left( T \right)$, the component volume fractions $\left( \phi_m \right)$, and the component internal energies $\left( e_m \right)$, from the following equations.
\begin{equation}
    p=p_m,\qquad T=T_m,\qquad \sum_{m=1}^M \phi_m = 1, \qquad \sum_{m=1}^M Y_m e_m = e.
\end{equation}
To achieve a stable equilibrium, it requires that all phases be present with non-negative volume fractions throughout the entire simulation domain.
This is achieved by initializing the problem with a minimum volume fraction (typically $\phi_{\text{min}}\lesssim10^{-6}$) and including additional criteria for volume fraction diffusion (Sections~\ref{sec:div} and \ref{sec:grad}) or mass fraction diffusion (Section~\ref{sec:LAD}) based on out-of-bounds values of volume fraction and/or mass fraction.
This equilibration method is stable in the well-mixed interface region, but can result in stability issues outside of the interface region, where the volume fraction of a material tends to become very small---a phenomenon exacerbated by high-order discretization methods.



\subsection{Localized artificial diffusivity \label{sec:LAD}}


LAD methods have long proven useful in conjunction with high-order compact derivative schemes to provide necessary solution-adaptive and localized diffusion to capture discontinuities and introduce a mechanism for subgrid dissipation.
Regardless of the choice of interface-capturing method, LAD is required in the momentum, energy, and kinematic equations, in all calculations, to provide necessary regularization.
For instance, the artificial shear viscosity, $\mu^*$, primarily serves as a subgrid dissipation model, whereas the artificial bulk viscosity, $\beta^*$, enables shock capturing, and the artificial thermal conductivity, $\kappa^*$, captures contact discontinuities.
The artificial kinematic diffusivities $\left(g^{e*} ~\text{and}~ g^{p*}\right)$ facilitate capturing of strain discontinuities, particularly in regions of sustained shearing.


When LAD is also used for interface regularization (to capture material interfaces), the artificial diffusivity of species $m$, $D^*_m$, is activated, in which the coefficient $C_D$ controls the interface diffusivity and the coefficient $C_Y$ controls the diffusivity when the mass fraction goes out of bounds. 
When using the volume-fraction-based approaches for interface regularization (Sections \ref{sec:div} and\ref{sec:grad}), it is often unnecessary to also include the species LAD $\left(D^*_m=0\right)$; however, the species LAD seems to be necessary for some problems in conjunction with these other interface regularization approaches.

The artificial diffusivities are described below, where the overbar denotes a truncated Gaussian filter applied along each grid direction; $\Delta_i$ is the grid spacing in the $i$ direction; $\Delta_{i,\mu}$, $\Delta_{i,\beta}$, $\Delta_{i,\kappa}$, $\Delta_{i,Y_m}$, and $\Delta_{i,g}$ are weighted grid length scales in direction $i$; $c_s$ is the linear longitudinal wave (sound) speed; $H$ is the Heaviside function; and $\varepsilon=10^{-32}$. 
\begin{equation}
    \mu^* = C_{\mu} \overline{\rho\left\lvert \sum_{k=1}^3 \frac{\partial^r S}{\partial x_k^r } \Delta^r_k \Delta^2_{k,\mu} \right\rvert}; \qquad \Delta_{i,\mu} = \Delta_i.
\end{equation} 
\begin{equation}
    \beta^* = C_{\beta} \overline{\rho f_{sw}\left\lvert \sum_{k=1}^3 \frac{\partial^r \left( \bm{\nabla}\cdot\bm{u} \right)}{\partial x_k^r } \Delta^r_k \Delta^2_{k,\beta} \right\rvert}; \qquad \Delta_{i,\beta} = \Delta_i \frac{\displaystyle \left( \frac{\partial\rho}{\partial x_i}\right)^2}{\displaystyle \sum_{k=1}^3 \left( \frac{\partial \rho}{\partial x_k}\right)^2 +\varepsilon}.
\end{equation} 
\begin{equation}
    \kappa^* = C_{\kappa} \overline{\frac{\rho c_s}{T}\left\lvert \sum_{k=1}^3 \frac{\partial^r e_h}{\partial x_k^r } \Delta^r_k \Delta_{k,\kappa} \right\rvert}; \qquad \Delta_{i,\kappa} = \Delta_i \frac{\displaystyle \left( \frac{\partial e_h}{\partial x_i}\right)^2}{ \displaystyle \sum_{k=1}^3 \left( \frac{\partial e_h}{\partial x_k}\right)^2 +\varepsilon}.
\end{equation} 
\begin{equation}
     \begin{aligned}
        D^*_m = \textrm{max}&\left\{ C_{D} \overline{c_s\left\lvert \sum_{k=1}^3 \frac{\partial^r Y_m}{\partial x_k^r } \Delta^r_k \Delta_{k,D} \right\rvert},\, C_Y \overline{\frac{c_s}{2} \left( \left\lvert Y_m \right\rvert -1 +\left\lvert 1-Y_m \right\rvert \right) \sum_{k=1}^3 \Delta_{k,Y}} \right\}; \\
        &\Delta_{i,D} = \Delta_i \frac{\displaystyle \left( \frac{\partial Y_m}{\partial x_i}\right)^2}{\displaystyle \sum_{k=1}^3 \left( \frac{\partial Y_m}{\partial x_k}\right)^2 +\varepsilon}; \qquad \Delta_{i,Y} = \Delta_i \frac{\displaystyle \left\lvert \frac{\partial Y_m}{\partial x_i}\right\rvert}{\sqrt{\displaystyle \sum_{k=1}^3 \left( \frac{\partial Y_m}{\partial x_k}\right)^2} +\varepsilon}.
    \end{aligned}
\end{equation} 
\begin{equation}
    g^* = C_g \overline{c_s\left\lvert \sum_{k=1}^3 \frac{\partial^r E^g}{\partial x_k^r } \Delta^r_k \Delta_{k,g} \right\rvert}; \qquad \Delta_{i,g} = \Delta_i \frac{\displaystyle \left( \frac{\partial E^g}{\partial x_i}\right)^2}{\displaystyle \sum_{k=1}^3 \left( \frac{\partial E^g}{\partial x_k}\right)^2 +\varepsilon}.
\end{equation}
\begin{equation}
    f_{sw} = H\left( -\bm{\nabla}\cdot\bm{u}\right) \frac{\left(\bm{\nabla}\cdot\bm{u}\right)^2}{\left(\bm{\nabla}\cdot\bm{u}\right)^2 + \left|\bm{\nabla}\times\bm{u} \right|^2 +\varepsilon}.
\end{equation}
Here, $S=\sqrt{S_{ij}S_{ij}}$ is a norm of the strain rate tensor, $S$, and $E^g = \sqrt{\frac{2}{3} E^g_{ij}E^g_{ji}}$ is a norm of the Almansi finite-strain tensor associated with the $g$ equations, $E^{g}_{ij} = \frac{1}{2}\left( \delta_{ij} - g_{ki} g_{kj} \right)$.
The artificial kinematic diffusivities, ${g^e}^*$ and ${g^p}^*$, are obtained by using the equation for ${g}^*$, but with the Almansi strain based on only the elastic or plastic component of the inverse deformation gradient tensor, respectively.
We observe that LAD is not strictly necessary to ensure stability for the $\textbf{g}^e$ equations; in fact, it has not been included in previous simulations \citep{ghaisas2018unified, subramaniam2018high}, because the elastic deformation is often small relative to the plastic deformation, but LAD is necessary to provide stability for the $\textbf{G}^p$ equations, especially when the interface is re-shocked, resulting in sharper gradients in the plastic deformation relative to the elastic deformation. 

Typical values for the model coefficients are $\zeta^e=\zeta^p=0.5$, $C_{\mu}=2\times10^{-3}$, $C_{\beta}=1$, $C_{\kappa}=1\times10^{-2}$, $C_{D}=3\times10^{-3}$, $C_{Y}=1\times10^{2}$, and $C_{g}=1$; these values are used in the subsequent simulations unless stated otherwise. 
However, these coefficients often need to be specifically tailored to the problem; for example, the bulk viscosity coefficient can be increased to more effectively capture strong shocks in materials with large stiffening pressures.

\subsection{Fully conservative divergence-form approach to interface regularization \label{sec:div}}

In this method, interface regularization is achieved with the use of diffusion and sharpening terms that balance each other.
This results in constant interface thickness during the simulation, unlike the LAD method, in which the interface thickness increases over time due to the absence of interface sharpening fluxes.
All regularization terms are constructed in divergence form, resulting in a method that conserves the mass of individual species as well as the mixture momentum and total energy.

Following \citet{jain2020conservative}, we consider the implied volume fraction transport equation for phase $m$, with the interface regularization volume fraction flux $\left(a_m\right)_k$,
\begin{equation}
    \frac{\partial \phi_m}{\partial t} + u_k \frac{\partial \phi_m}{\partial x_k} = \frac{\partial \left( a_m \right)_k}{\partial x_k}.
\end{equation}
In this work, this equation is not directly solved, because the volume fraction is closed during the pressure and temperature equilibration process (Section~\ref{sec:equilibrium}), but the action of this volume fraction flux is consistently incorporated into the system through the coupling terms with the other governing equations.
We employ the coupling terms proposed by \citet{jain2020conservative} for the mass, momentum, and energy equations, and propose new consistent coupling terms for the kinematic equations.

Using the relationship of density $\left(\rho\right)$ and mass fraction $\left(Y_m\right)$ to component density $\left(\rho_m\right)$ and volume fraction $\left(\phi_m\right)$ for material $m$ ($\rho Y_m = \rho_m \phi_m$, with no sum on repeated $m$), we can describe the interface regularization term for each material mass transport equation,
\begin{equation}
    J_m = \frac{\partial \left(a_m\right)_k \rho_m}{\partial x_k}, \qquad \text{with no sum on repeated }m.
\end{equation}
Consistent regularization terms for the momentum and energy equations follow,
\begin{equation}
    F_i = \frac{\partial \left(a_m\right)_k \rho_m u_i}{\partial  x_k} \qquad\text{and}\qquad H = \frac{\partial}{\partial x_k}\left\{ \left(a_m\right)_k \left[\frac{1}{2} \rho_m u_j u_j  + \left(\rho h\right)_m \right] \right\},
\end{equation}
 in which the enthalpy of species $m$ is described,
\begin{equation}
    \left(\rho h\right)_m = e_m \rho_m +p_m, \qquad\textrm{with no sum on repeated }m.
\end{equation}
Consistent regularization terms for the kinematic equations take the form

\begin{equation}
    K^e_{ij} = \frac{1}{3} \frac{1}{\rho} g^e_{ij} \sum_m J_m
\end{equation}
This is derived by considering a transport equation for $\det \textbf{g}^e$:
\begin{equation}
    \DDt{}(\det{\textbf{g}^e}) = \pp{\det{\textbf{g}^e}}{\textbf{g}^e} : \DDt{}{\textbf{g}^e}
\end{equation}
where $\DDt{}$ denotes the material derivative, and ``$:$" denotes the tensor inner product. Using identities from tensor calculus and the properties of the multiplicative decomposition, this can be simplified to
\begin{equation}
    \DDt{\rho} = \rho (\textbf{g}^e)^{-T} : \DDt{}{\textbf{g}^e}
\end{equation}
Plugging in the transport equations for $\textbf{g}^e$ and $\rho$, converting to index notation, and ignoring the other artificial terms, this becomes
\begin{equation}
    - \rho \pp{u_k}{x_k} + \sum_m J_m  = \rho (g^e)^{-1}_{ji} \Big( -g^e_{ik} \pp{u_k}{x_j} + K^e_{ij} \Big)
\end{equation}
The terms involving velocity cancel, leaving
\begin{equation}
    \sum_m J_m  = \rho \Big( (g^e)^{-1}_{ji} K^e_{ij} \Big)
\end{equation}
This relationship can be satisfied in many ways, but the form employed here is chosen for simplicity. 
No sharpening term is required in the equations for plastic deformation because in the multiplicative decomposition, volume change is entirely described by $\textbf{g}^e$. 

The volume fraction interface regularization flux for phase $m$ is described by
\begin{equation}  \label{eqn:div_vfr}
\left(a_m\right)_k = \Gamma \left[\underbrace{\epsilon\frac{\partial \phi_m}{\partial x_k}}_{\substack{\textrm{interface} \\ \textrm{diffusion}}} -  \underbrace{s_m \left( \hat{n}_m \right)_k}_{\substack{\textrm{interface} \\ \textrm{sharpening}}}\right] L_m + \underbrace{\Gamma^* \epsilon D_b \frac{\partial \phi_m}{\partial x_k}}_{\substack{\textrm{out-of-bounds} \\ \textrm{diffusion}}}, \qquad \text{with no sum on repeated }m,
\end{equation}
with the interface sharpening term
\begin{equation}
s_m = \begin{cases}
        \left(\phi_m -\phi_{m}^{\epsilon}\right) \left( 1 -\sum_{\substack{n=1 \\ n\neq m}}^M \phi_n^{\epsilon}- \phi_m \right), & \text{for } \phi_{m}^{\epsilon} \leq \phi_m \leq 1 - \sum_{\substack{n=1 \\ n\neq m}}^M \phi_n^{\epsilon} \\
        0, & \text{else }
      \end{cases},
\end{equation}
in which $\phi_{m}^{\epsilon}$ denotes the minimum allowable volume fraction for phase $m$; this floor promotes physically realizable solutions to the pressure and temperature equilibria, which would otherwise not be well behaved if the mass or volume fraction exceeded the physically realizable bounds between zero and one.
We assume $\phi_{m}^{\epsilon}=1\times10^{-6}$ unless stated otherwise.
The optional mask term,
\begin{equation}
L_m = \begin{cases}
        1, & \text{for } \phi_{m}^{\epsilon} \leq \phi_m \leq \sum_{\substack{n=1 \\ n\neq m}}^M \phi_n^{\epsilon} \\
        0, & \text{else }
      \end{cases},
\end{equation}
localizes the interface diffusion and interface sharpening terms to the interface region, restricting the application of the non-compactly discretized terms to the interface region.
Unlike the gradient-form approach, this mask in the divergence-form approach is not necessary for stability, as demonstrated by \citet{jain2020conservative}.
The interface normal vector for phase $m$ is given by
\begin{equation} 
\left( \hat{n}_m \right)_k = \frac{\partial \phi_m}{\partial x_k} / \left\lvert \frac{\partial \phi_m}{\partial x_i}\right\rvert, \qquad \left\lvert \frac{\partial \phi_m}{\partial x_i}\right\rvert = \sqrt{\frac{\partial \phi_m}{\partial x_i}\frac{\partial \phi_m}{\partial x_i}}, \qquad\textrm{with no sum on repeated }m.
\end{equation}
The out-of-bounds diffusivity, described by 
\begin{equation}
    D_b = \max_m\overline{\left[1-\phi_m/\left(\phi_m^{\epsilon}\right)^b\right]}_{\textrm{no sum in }m} (1 - L_m),\qquad b=\frac{1}{2},
\end{equation}
maintains $\phi_m\gtrsim\phi_m^{\epsilon}$.
The overbar denotes the same filtering operation as applied to the LAD diffusivities.
A user-specified length scale, $\epsilon\approx \Delta x$, typically on the order of the grid spacing, controls the equilibrium thickness of the diffuse interface.
The velocity scale,
$\Gamma\approx u_{max}$ 
controls the timescale over which the interface diffusion and interface sharpening terms drive the interface thickness to equilibrium.
The velocity scale for the out-of-bounds volume fraction diffusivity is also specified by the user, with $\Gamma^*\gtrsim \Gamma$.
Volume fraction compatibility is enforced by requiring that $\sum_{m=1}^M \left(a_m\right)_k = 0$.

\subsection{Quasi-conservative gradient-form approach to interface regularization \label{sec:grad}}

As with the divergence-form approach, the interface regularization in this approach is achieved with the use of diffusion and sharpening terms that balance each other.
Therefore, this method also results in constant interface thickness during the simulation.
\citet{shukla2010interface} discuss disadvantages associated with the divergence-form approach due to the numerical differentiation of the interface normal vector.
The numerical error of these terms can lead to interface distortion and grid imprinting due to the anisotropy of the derivative scheme.
Ideally, we would like to have a regularization method that is conservative and that does not require any numerical differentiation of the interface normal vector.
However, starting with the assumption of conservation, for nonzero regularization flux, we see that numerical differentiation of the interface normal vector can only be avoided in the limit that the divergence of the interface normal vector goes to zero.
This limit corresponds to the limit of zero interface curvature, which cannot be avoided in multidimensional problems.
Therefore, this illustrates that a conservative method cannot be constructed for multidimensional applications without requiring differentiation of the interface normal vector; the non-conservative property (undesirable) of the gradient-form approach is a necessary consequence of the circumvention of interface-normal differentiation (desirable).
This is demonstrated below, in which the phase subscript has been dropped.
\begin{equation}
    \begin{gathered}
        \frac{\partial }{\partial x_k} \left\{ \Gamma \left[\epsilon \frac{\partial \phi}{\partial x_k}  + \phi \left( 1-\phi\right) \hat{n}_k \right]\right\}\\ 
        = \hat{n}_k \frac{\partial }{\partial x_k} \left\{ \Gamma \epsilon \left\lvert\frac{\partial \phi}{\partial x_j}\right\rvert \right\}  + \left\{ \Gamma \epsilon \left\lvert\frac{\partial \phi}{\partial x_j}\right\rvert \right\} \frac{\partial \hat{n}_k }{\partial x_k} + \frac{\partial \Gamma \phi \left( 1-\phi\right)}{\partial x_k} \hat{n}_k  + \Gamma \phi \left( 1-\phi\right)  \frac{\partial \hat{n}_k}{\partial x_k} \\
        \stackrel[\nabla\cdot\Vec{\bm{n}}\rightarrow 0]{\longrightarrow}{}  \hat{n}_k \frac{\partial }{\partial x_k} \left\{ \Gamma \left[\epsilon \left\lvert\frac{\partial \phi}{\partial x_j}\right\rvert + \phi\left( 1-\phi\right)\right]\right\},\\
    \end{gathered}
\end{equation}
in which the final expression is obtained in the limit of $ \nabla \cdot \vec{\bm{n}} \rightarrow 0$.

Following \citet{shukla2010interface}, we arrive at an implied volume fraction transport equation for phase $m$, with the interface regularization volume fraction term $\alpha_m$,
\begin{equation}
    \frac{\partial \phi_m}{\partial t} + u_k \frac{\partial \phi_m}{\partial x_k} = \left( n_m \right)_k \frac{\partial \alpha_m}{\partial x_k}.
    \label{eq:grad_vol}
\end{equation}
Unlike the divergence-form approach, the gradient-form approach requires no numerical differentiation of interface normal vectors, but it consequently results in conservation error.
Like the divergence-form approach, this volume fraction transport equation is not directly solved, because the volume fraction is closed during the pressure and temperature equilibration process (Section~\ref{sec:equilibrium}), but the action of the volume fraction regularization term is consistently incorporated into the system of equations for mass, momentum, energy, and kinematic quantities through quasi-conservative coupling terms.

We employ an interface regularization term for each component mass transport equation consistent with the interface regularization volume fraction term,
\begin{equation}
    J_m = \left( n_m \right)_k \frac{\partial \alpha_m \rho_m}{\partial x_k}, \qquad \textrm{with no sum on repeated }m.
\end{equation}
Because of the assumption of pressure and temperature equilibrium (volume fraction is a derived variable---not an independent state variable), it is important to form mass transport regularization terms consistently with the desired volume fraction regularization terms.
In the method of \citet{tiwari2013diffuse}, the terms do not need to be fully consistent (e.g., the component density is assumed to be slowly varying); the terms only need to produce similar interface profiles in the limit of $\Gamma\rightarrow\infty$ \citep{shukla2010interface}, because the volume fraction is an independent state variable.
Following the assumption of \citet{tiwari2013diffuse} that the velocity, specific energy, and kinematic variables (but not the mixture density) vary slowly across the interface, the stability of the method is improved by further relaxing conservation of the coupled equations.
For example, the consistent regularization term for the momentum equation reduces to
\begin{equation}
    F_i = \sum_m \left( n_m \right)_k \frac{\partial \alpha_m \rho_m u_i}{\partial x_k} \approx \sum_m \left( n_m \right)_k \frac{\partial \alpha_m \rho_m}{\partial x_k} u_i.
\end{equation}
Similarly, the consistent regularization term for the energy equation reduces to
\begin{equation}
    H = \sum_m \left( n_m \right)_k \frac{\partial }{\partial x_k} \left[ \alpha_m \rho_m \left(\frac{1}{2} u_j u_j  + h_m \right) \right] \approx \sum_m \left( n_m \right)_k \frac{\partial \alpha_m \rho_m }{\partial x_k} \left(\frac{1}{2} u_j u_j  + h_m \right).
\end{equation}
As with the divergence method, consistent regularization terms for the kinematic equations take the form
\begin{equation}
    K^e_{ij} = \frac{1}{3} \frac{1}{\rho} g^e_{ij} \sum_m J_m.
\end{equation}
No sharpening terms are required for the plastic deformation.

The volume fraction interface regularization flux for phase $m$ is defined by
\begin{equation} \label{eqn:grad_vfr}
\alpha_m = \Gamma \left(\underbrace{\epsilon\left\lvert \frac{\partial \phi_m}{\partial x_i}\right\rvert}_{\substack{\textrm{interface} \\ \textrm{diffusion}}} -  \underbrace{s_m}_{\substack{\textrm{interface} \\ \textrm{sharpening}}} \right) \mathscr{L}_m, \qquad \textrm{with no sum on repeated }m.
\end{equation}
The volume fraction out-of-bounds diffusion term employed in the divergence-form approach (Eq. \ref{eqn:div_vfr}) is also active in the gradient-form approach. 
The gradient-form discretization of this term (including an equivalent volume fraction out-of-bounds term in Eq. \ref{eqn:grad_vfr}) exhibits poor stability away from the interface, whereas the divergence-form approach does not.
Following \citet{shukla2010interface} and \citet{tiwari2013diffuse}, a necessary mask term blends the interface regularization terms to zero as the volume fraction approaches the specified minimum or maximum, thereby avoiding instability of the method away from the interface, where the calculation of the surface normal vector may behave spuriously and lead to compounding conservation error,
\begin{equation}
\mathscr{L}_m = \begin{cases}
        \tanh \left[ \left( \displaystyle \frac{s_m}{\phi_m^{\mathscr{L}}}\right)^2 \right], & \text{for } \phi_{m}^{\epsilon} \leq \phi_m \leq 1 - \sum_{\substack{n=1 \\ n\neq m}}^M \phi_n^{\epsilon} \\
        0, & \text{else }
      \end{cases},
\end{equation}
in which $\phi_m^{\mathscr{L}}\approx 1\times10^{-2}$ is a user-specified value controlling the mask blending function.
Other variables are the same as defined in the context of the divergence-form approach.

\subsection{Numerical method\label{sec:numerics}} 
The equations are discretized on an Eulerian Cartesian grid.
Time advancement is achieved using a five-stage, fourth-order, Runge-Kutta method, with an adaptive time step based on a Courant–Friedrichs–Lewy (CFL) condition.
Other than the interface regularization terms for the divergence-form approach, all spatial derivatives are computed using a high-resolution, penta-diagonal, tenth-order, compact finite-difference scheme described by \cite{lele1992compact}.  
This scheme is applied in the domain interior and near the boundaries in the cases of symmetry, anti-symmetry, or periodic boundary conditions.
Otherwise, boundary derivatives are reduced to a fourth-order, one-sided, compact difference scheme.

The interface sharpening and interface diffusion regularization terms in the divergence-form approach are discretized using node-centered
derivatives, for which the fluxes to be differentiated are formed at the faces (staggered locations); linear terms (e.g., $\phi_{i}$) are interpolated from the nodes to the faces, where the nonlinear terms are formed [e.g., $\phi_{i+1/2}\left(1-\phi_{i+1/2}\right)$].
Here, we refer to the finite-difference grid points as nodes.
All variables are stored at the nodes (collocated).
If the nonlinear fluxes are not formed at the faces, poor stability is observed for node-centered finite-difference schemes of both compact and non-compact varieties due to the nonlinear interface sharpening term (see Appendix A).
A second-order scheme is used for discretization of the interface regularization terms throughout this work, with an exception in Section \ref{sec:bubble_advection} where both second-order and sixth-order (non-compact) discretization schemes are examined for these terms.
The second-order scheme recovers the finite-volume approach employed by \citet{jain2020conservative}, whereas the higher-order scheme provides increased resolution and formal accuracy; however, discrete conservation is not guaranteed. The sixth order explicit scheme used to compute first derivatives from nodes to faces or vice versa is
\begin{equation}
    f'_i = \frac{9}{384} \frac{f_{i+5/2}-f_{i-5/2}}{5h} - \frac{25}{128} \frac{f_{i+3/2}-f_{i-3/2}}{3h} + \frac{225}{192} \frac{f_{i+1/2}-f_{i-1/2}}{h}
\end{equation}
The sixth order interpolation scheme used for node to face or vice versa is
\begin{equation}
    \hat{f}_i =  \frac{3}{256}(f_{i+5/2}+f_{i-5/2}) - \frac{25}{256}(f_{i+3/2}+f_{i-3/2}) + \frac{75}{128}(f_{i+1/2}+f_{i-1/2})
\end{equation}

\noindent The out-of-bounds diffusion is discretized using the tenth-order pentadiagonal scheme for all interface regularization approaches.

A spatial dealiasing filter is applied after each stage of the Runge-Kutta algorithm to each of the conservative and kinematic variables to remove the top $10\%$ of the grid-resolvable wavenumber content, thereby mitigating against aliasing errors and numerical instability in the high-wavenumber range, which is not accurately resolved by the spatial derivative scheme.
The filter is computed using a high-resolution, penta-diagonal, eighth-order, compact Pad{\'e} filter, with cutoff parameters described by \cite{ghaisas2018unified}.

\section{Results}\label{sec:results}

In this section, we present the simulation results and evaluate the performance of the methods using classical test cases, such as: (a) advection of an air bubble in water, (b) shock interaction with a helium bubble in air, (c) shock interaction and the collapse of an air bubble in water, and (d) Richtmyer-Meshkov instability (RMI) of a copper-aluminium interface. The simulation test cases in the present study were carefully selected to assess: (1) the conservation property of the method; (2) the accuracy of the method in maintaining the interface shape; and (3) the ability of the method in maintaining constant interface thickness throughout the simulation.

Some of these test cases have been extensively studied in the past and have been used to evaluate the performance of various interface-capturing and interface-tracking methods.
Many studies look at these test cases to evaluate the performance of the methods in the limit of very fine grid resolutions.
For example, a typical value of the grid size is on the order of hundreds of mesh points across the diameter of a single bubble/droplet.
However, for practical application of these methods in the large-scale simulations of engineering interest\textemdash where there are thousands of droplets, e.g., in an atomization process\textemdash it is rarely affordable to use such fine grids to resolve a single droplet/bubble.
Therefore, in this study, we examine these methods in the opposite limit of relatively coarse grid resolution.
This limit is more informative of the true performance of these methods for practical applications.
All three diffuse-interface capturing methods are implemented in the Pad\'eOps solver \citep{padeops} 
to facilitate fair comparison of the methods with the same underlying numerical methods, thereby eliminating any solver/implementation-related bias in the comparison. 



The first test case (Section \ref{sec:bubble_advection}) is the advection of an air bubble in water.
This test case is chosen to evaluate the ability of the interface-capturing method to maintain the interface shape for long-time numerical integration and to examine the robustness of the method for high-density-ratio interfaces.
It is known that the error in evaluating the interface normal accumulates over time and results in artificial alignment of the interface along the grid \citep{chiodi2017reformulation, tiwari2013diffuse}.
This behavior is examined for each of the three methods.
The second test case (Section \ref{sec:helium_bubble}) is the shock interaction with a helium bubble in air.
This test case is chosen to evaluate the ability of the methods to conserve mass, to maintain constant interface thickness throughout the simulation, and to examine the behavior of the under-resolved features captured by the methods.
The third test case (Section \ref{sec:bubble_collapse}) is the shock interaction with an air bubble in water.
This test case is chosen to evaluate the robustness of the method to handle strong-shock/high-density-ratio interface interactions.
The fourth test case (Section \ref{sec:RMI}) is the RMI of a copper--aluminum interface.
This test case is chosen to illustrate the applicability of the methods to simulate interfaces between solid materials with strength, to examine the conservation properties of the methods in the limit of high interfacial curvature, to examine the ability of the methods to maintain constant interface thickness, and to assess the behavior of the under-resolved features captured by the methods.

For all the problems in this work, the mass fractions are initialized using the relations $Y_1 = \phi_1 \rho_1/\rho$ and $Y_2 = 1-Y_1$.  To evaluate the mass-conservation property of a method, the total mass, $m_k$, of the phase $k$ is calculated as
\begin{equation}
    m_k = \int_\Omega \rho Y_k dv, 
    \label{equ:mass}
\end{equation}
where the integral is computed over the computational domain $\Omega$.
To evaluate the ability of a method to maintain constant interface thickness, we define a new parameter---the interface-thickness indicator ($l$)---as
\begin{equation}
    l = \left(\frac{1}{\hat{n}\cdot \Vec{\nabla} \phi}\right),
    \label{equ:l}
\end{equation}
and compute the maximum and average interface thicknesses in the domain, using
\begin{equation}
    l_{max}= \max_{0.45 \leq \phi \leq 0.55}\left( l \right), \hspace{1cm} l_{avg}= \left\langle l \right\rangle_{0.45 \leq \phi \leq 0.55}.
    \label{equ:thick}
\end{equation}
respectively, where $\langle\cdot\rangle$ denotes an averaging operation. The range for $\phi$ is used to ensure that the interface thickness is evaluated around the $\phi=0.5$ isocontour because the quantity $l$ is most accurate in this region and goes to $\infty$ as $\phi\rightarrow 0,1$. 
Note that, occasionally, $l$ can become very large, within the region $0.45 \leq \phi \leq 0.55$, when there is breakup due to the presence of a saddle point in the $\phi$ field at the location of rupture. These unphysical values of $l$ show up in the computed $l_{max}$ values and are removed during the post-processing step by plotting a moving average of 5 local minima of $l_{max}$. The unphysical values of $l$, on the other hand, have only a small effect on the computed $l_{avg}$ values due to the averaging procedure.


\subsection{Advection of an air bubble in water}\label{sec:bubble_advection}


This section examines advection of a circular air bubble in water. A one-dimensional version of this test case has been extensively studied before, and has been previously used as a test of robustness of the method using various diffuse-interface methods in \citet{saurel1999simple,allaire2002five,murrone2005five,johnsen2012preventing,saurel2009simple,johnsen2006implementation,coralic2014finite,beig2015maintaining,capuano2018simulations}, and using a THINC method in \citet{shyue2014eulerian}. A two-dimensional advection of a bubble/drop has also been studied using a weighted-essentially non-oscillatory (WENO) and targeted-essentially non-oscillatory (TENO) schemes in \citet{haimovich2017numerical}.

In the current study, this test case is used to evaluate the ability of the methods in maintaining interface-shape for long-time integrations and as a test of robustness of the methods for high-density-ratio interfaces. Both phases are initialized with a uniform advection velocity. The problem domain spans $\left(0\leq x \leq 1;\, 0\leq y\leq 1\right)$, with periodic boundary conditions in both dimensions.
The domain is discretized on a uniform Cartesian grid of size $N_x = 100$ and $N_y = 100$.
The bubble has a radius of $25/89$ and is initially placed at the center of the domain. The material properties for the water medium used in this test case are $\gamma_1=4.4$, $\rho_1= 1.0$, ${p_{\infty}}_1=6\times10^3$, $\mu_1 = 0$, and ${\sigma_Y}_1 = 0$.
The material properties for the air medium used in this test case are $\gamma_2=1.4$, $\rho_2= 1\times10^{-3}$, ${p_{\infty}}_2 = 0$, $\mu_2 = 0$, and ${\sigma_Y}_2 = 0$, where $\gamma_k,\rho_k,{p_{\infty}}_k,\mu_k$, and ${\sigma_Y}_k$ are the ratio of specific heats, density, stiffening pressure, shear modulus, and yield stress of phase $k$, respectively.

The initial conditions for the velocity, pressure, volume fraction, and density are 
\begin{equation}
    u=5, \quad
    v=0, \quad
    p=1, \quad
    \phi_1 = \phi_1^{\epsilon} + \left(1-2\phi_1^{\epsilon}\right) f_{\phi}, \quad
    \phi_2 = 1-\phi_1, \quad \rho = \phi_1 \rho_1 + \phi_2 \rho_2,
\end{equation}
respectively, in which the volume fraction function, $f_{\phi}$, is given by
\begin{equation}
    f_{\phi} =\frac{1}{2} \left\{1 -\textrm{erf} \left[\frac{625/7921-\left(x-1/2\right)^2 - \left(y-1/2\right)^2}{3\Delta x}\right] \right\}. 
\end{equation}
For this problem, the interface regularization length scale and the out-of-bounds velocity scale are defined by $\epsilon=\Delta x = 1.0\times10^{-2}$ and $\Gamma^*=5.0$, respectively.

\begin{figure}
    \centering
    \includegraphics[width=\textwidth]{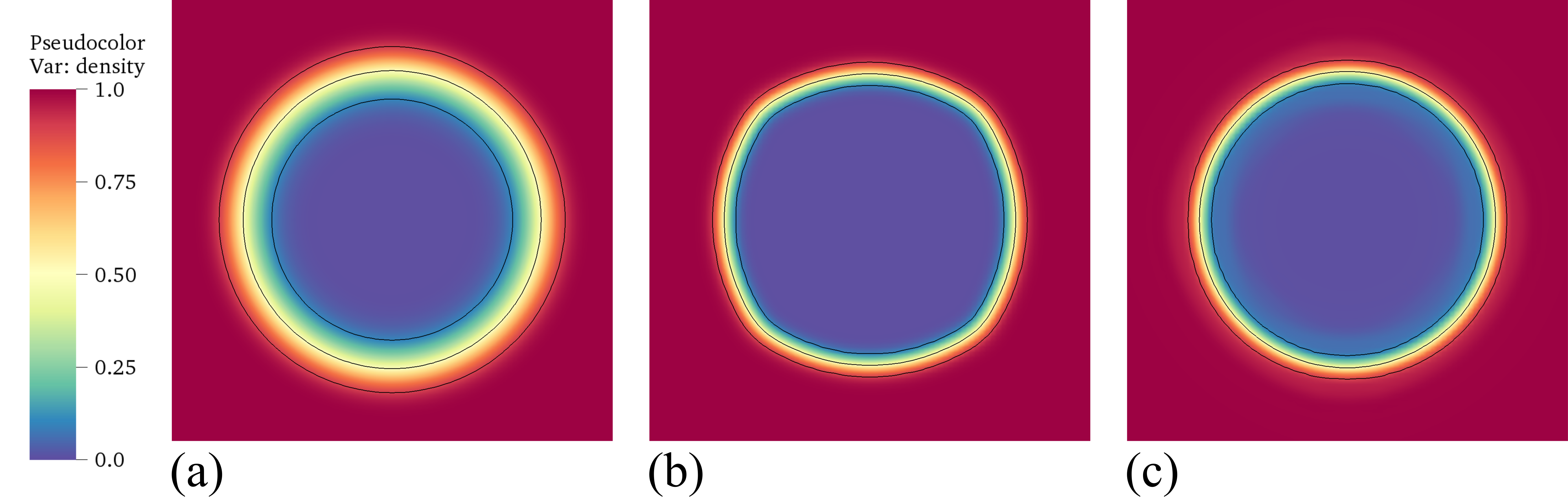}
    \caption{Comparison of the final state of the bubble after five flow-through times using (a) LAD approach, (b) divergence-form approach, and (c) gradient-form approach. The three solid black lines denote the isocontours of the volume fraction values of 0.1, 0.5, and 0.9, representing the interface region.}
    \label{fig:bubble_advection}
\end{figure}

The simulation is integrated for a total physical time of $t=1$ units, and the bubble at this final time is shown in Figure \ref{fig:bubble_advection}, facilitating comparison among the LAD, divergence-form, and gradient-form methods.
All three methods perform well and are stable for this high-density-ratio case.
The consistent regularization terms included in the momentum and energy equations are necessary to maintain stability.
The divergence-form approach results in relatively faster shape distortion compared to the LAD and gradient-form approaches.
This shape distortion is due to the accumulation of error resulting from numerical differentiation of the interface normal vector, which is required in the divergence-form approach but not the other approaches.
A similar behavior of interface distortion was seen when the velocity was halved and the total time of integration was doubled, thereby confirming that this behavior is reproducible for a given flow-through time (results not shown). 
\begin{figure}
    \centering
    \includegraphics[width=0.75\textwidth]{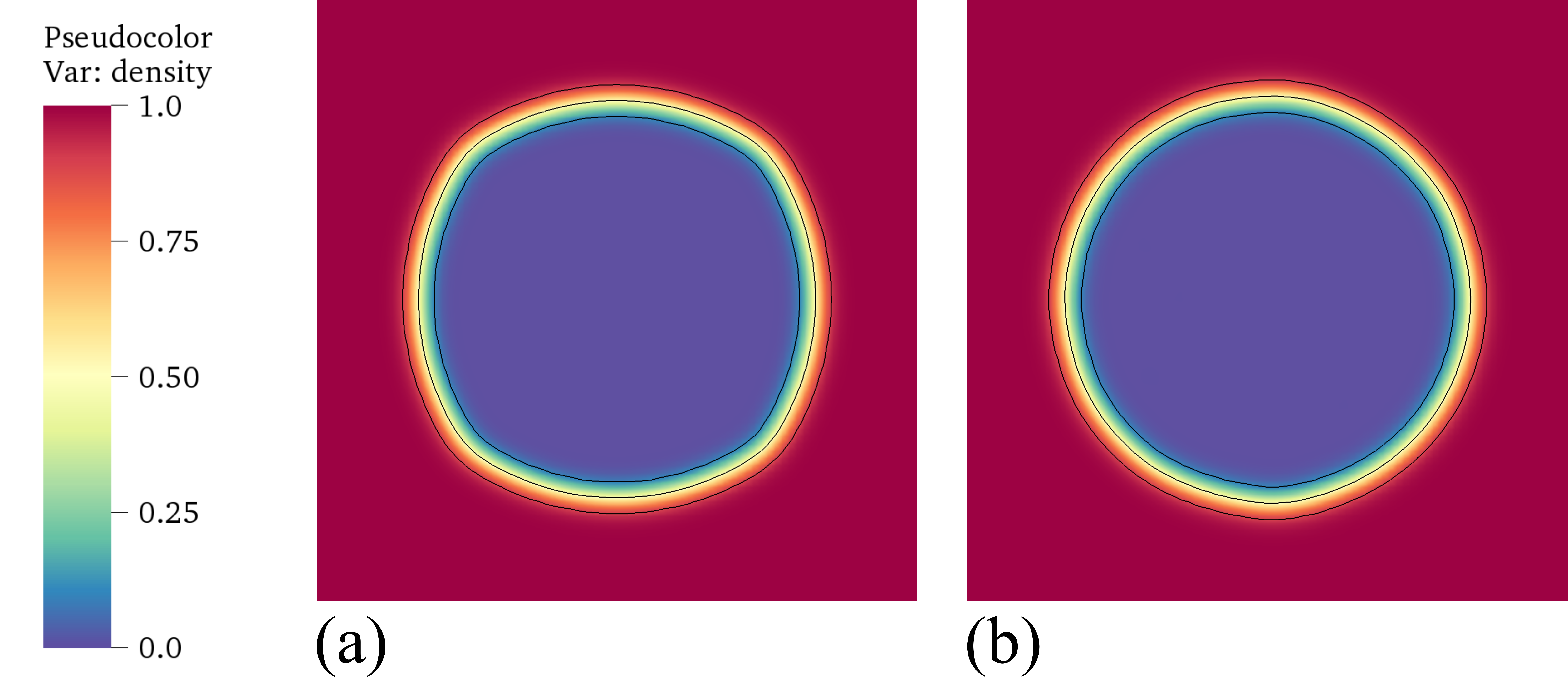}
    \caption{Comparison of the  state of the bubble after five flow-through times using the divergence-form approach with (a) second-order scheme and (b) sixth-order scheme. The three solid black lines denote the isocontours of the volume fraction values of 0.1, 0.5, and 0.9, representing the interface region.} 
    \label{fig:bubble_advection_high}
\end{figure}

%
%
%
%
%
%

Two possible ways to mitigate the interface distortion are by refining the grid or by using a higher-order scheme for the interface-regularization terms.
Because we are interested in the limit of coarse grid resolution, we study the effect of using an explicit sixth-order finite-difference scheme to discretize the interface regularization terms.
As described in Section 2.8, finite-difference schemes may be used to discretize the interface regularization terms\textemdash without resulting in spurious behavior\textemdash if the nonlinear interface sharpening and the counteracting diffusion terms are formed at the grid faces (staggered locations), from which the derivatives at the grid points (nodes) may be calculated. 
Comparing the second-order and sixth-order schemes for the interface regularization terms of the divergence-form approach, the final state of the advecting bubble is shown in Figure \ref{fig:bubble_advection_high}.
The interface distortion is significantly reduced using the sixth-order scheme.


Since the focus of the current work is on the evaluation of methods in a relatively coarser grid, we repeat the simulation of advection of an air bubble in water by scaling down the problem (in length and time) by a factor of $10$ without changing the number of grid points. The domain length is kept the same, but the new bubble radius is $2.5/89$, and the simulation is integrated for a total physical time of $t=0.1$ units. At this resolution, the bubble has $\approx5$ grid points across its diameter, which represents a more realistic scenario that is encountered in large-scale engineering simulations.

\begin{figure}
    \centering
    \includegraphics[width=\textwidth]{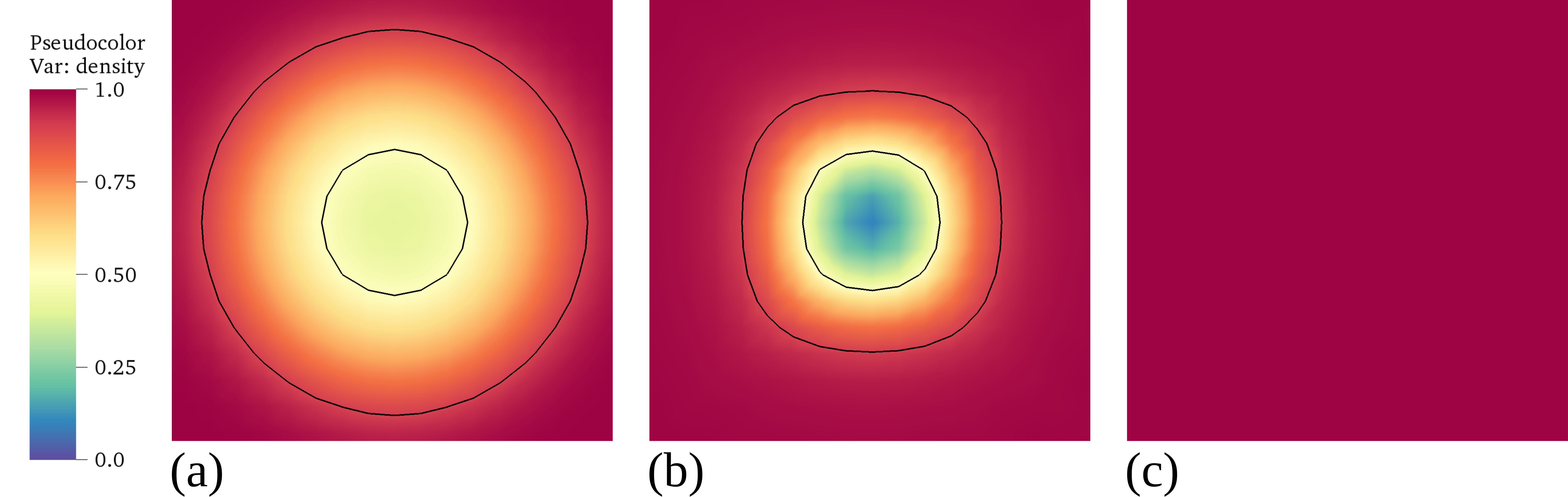}
    \caption{Comparison of the final state of the bubble on a coarse grid after five flow-through times using (a) LAD approach, (b) divergence-form approach, and (c) gradient-form approach. The two solid black lines denote the isocontours of the volume fraction values of 0.5 and 0.9, representing the interface region.}
    \label{fig:bubble_advection_small}
\end{figure}

\begin{figure}
    \centering
    \includegraphics[width=0.5\textwidth]{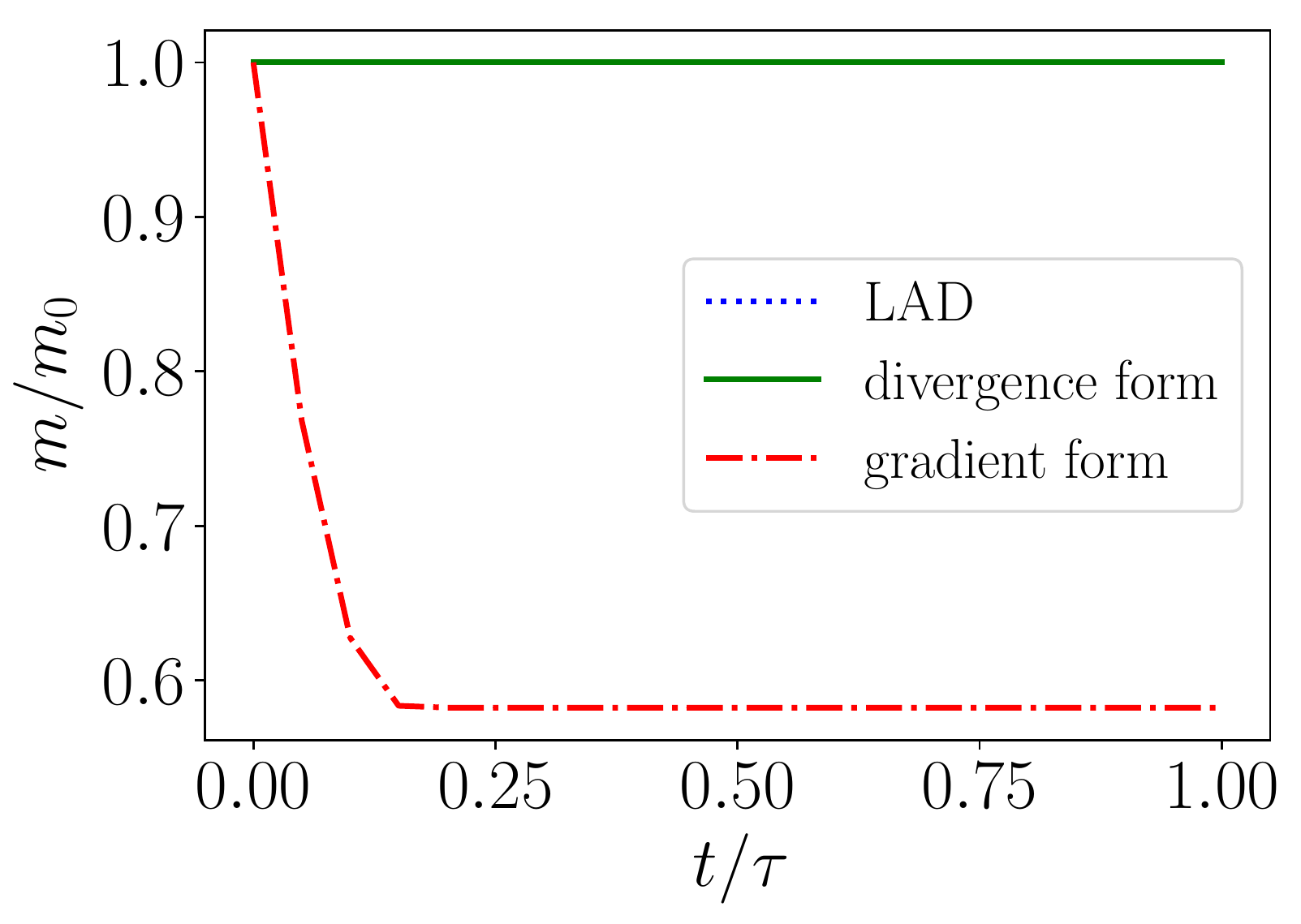}
    \caption{Plot of total mass, $m$, of the bubble by various methods, where $m_0$ is the mass at time $t=0$.}
    \label{fig:bubble_advection_mass}
\end{figure}

The bubble at the final time is shown in Figure \ref{fig:bubble_advection_small}, for all the three methods. Similar to the more refined case above with the second-order finite-volume scheme, the divergence-form approach results in relatively faster shape distortion compared to the LAD method. Whereas, the gradient-form approach results in apparent complete loss of the bubble. Comparing this result with the refined simulation in Figure \ref{fig:bubble_advection}, this observation of mass loss on coarse grids is in good agreement with our hypothesis that the conservation error is proprotional to the local interface curvature and the under-resolved features are more prone to being lost due to the non-conservative nature of the method. This makes the gradient-form approach unsuitable for large-scale engineering simulations where it is only possible to afford a couple of grids points across a bubble/drop. To quantify the amount of mass loss with the gradient-form approach, the bubble mass is plotted against time for all three methods in Figure \ref{fig:bubble_advection_mass}. Interestingly, around $40\%$ of the bubble mass is lost during the early time in the simulation and then the bubble mass saturates. This is due to the traces of mass of bubble that is still present in the domain, that is not under-resolved, after all the fine features are lost due to the conservation error.



\subsection{Shock interaction with a helium bubble in air \label{sec:helium_bubble}}


This section examines the classic test case of a shock wave traveling through air followed by an interaction with a stationary helium bubble. This case has been extensively studied using various numerical methods and models, such as a front-tracking method in \citet{terashima2009front}; an arbitrary-Lagrangian Eulerian (ALE) method in \citet{daude2014numerical}; anti-diffusion interface-capturing method in \citet{so2012anti}; a ghost-fluid method (GFM) in \citet{fedkiw1999non,bai2017sharp}; a LAD diffuse-interface approach in \citet{cook2009enthalpy}; a gradient-form diffuse-interface approach in \citet{shukla2010interface}, and other diffuse-interface methods that implicitly capture the interface (no explicit interface-capturing method) using a WENO scheme in \citet{johnsen2006implementation,coralic2014finite}, using a TENO scheme in \citet{haimovich2017numerical}, and using a WCNS scheme in \citet{wong2017high}. This test case has also been simulated with an adaptive-mesh-refinement technique in \citet{quirk1996dynamics} where a refined grid is used around the interface to improve the accuracy. More recently, this test case has also been studied in a three-dimensional setting in \citet{deng2018high}.

To examine the interface regularization methods, we model this problem without physical species diffusion; therefore, the interface regularization methods for immiscible phases are applicable, because no physical molecular mixing should be exhibited by the underlying numerical model.
The use of immiscible interface-capturing methods to model the interface between the gases in this problem is also motivated by the experiments of \citet{haas1987interaction}.
In these experiments, the authors use a thin plastic membrane to prevent molecular mixing of helium and air.  

The problem domain spans $\left(-2\leq x \leq 4;\, 0\leq y\leq 1\right)$, with periodic boundary conditions in the $y$ direction.
A symmetry boundary is applied at $x=4$, representing a perfectly reflecting wall, and a sponge boundary condition is applied over $\left(-2\leq x \leq -1.5\right)$, modeling a non-reflecting free boundary. The problem is discretized on a uniform Cartesian grid of size $N_x = 600$ and $N_y = 100$.
The bubble has a radius of $25/89$ and is initially placed at the location $(x=0,y=1/2)$.
The material properties for the air medium are described by $\gamma_1=1.4$, $\rho_1= 1.0$, ${p_{\infty}}_1=0$, $\mu_1 = 0$, and ${\sigma_Y}_1 = 0$.
The material properties for the helium medium are described by $\gamma_2=1.67$, $\rho_2= 0.138$, ${p_{\infty}}_2 = 0$, $\mu_2 = 0$, and ${\sigma_Y}_2 = 0$.

The initial conditions for the velocity, pressure, volume fraction, and density are
\begin{equation}
    \begin{gathered}
        u=u^{(2)} f_s + u^{(1)} \left(1-f_s\right), \quad
        v=0, \quad
        p=p^{(2)} f_s + p^{(1)} \left(1-f_s\right), \\
        \phi_1 = \phi_1^{\epsilon} + \left(1-2\phi_1^{\epsilon}\right) f_{\phi}, \quad
        \phi_2 = 1-\phi_1, \quad
        \rho = \left(\phi_1 \rho_1 + \phi_2 \rho_2\right) \left[ \rho^{(2)}/\rho^{(1)} f_s  +  \left(1-f_s\right) \right],
        \end{gathered}
\end{equation}
respectively, in which the volume fraction function, $f_{\phi}$, and the shock function, $f_s$, are given by
\begin{equation}
    f_{\phi} =\frac{1}{2} \left\{1 -\textrm{erf} \left[\frac{625/7921-x^2 - \left(y-1/2\right)^2}{\Delta x}\right] \right\}\quad \textrm{and} \quad f_s = \frac{1}{2} \left[ 1 - \textrm{erf}\left(\frac{x+1}{2 \Delta x}\right) \right],
\end{equation}
respectively, with jump conditions across the shock for velocity $\left(u^{(1)}=0;\,u^{(2)}=0.39473\right)$, density $\left(\rho^{(1)}=1\right.$, $\left.\rho^{(2)}=1.3764\right)$, and pressure $\left(p^{(1)}=1;\,p^{(2)}=1.5698\right)$.
For this problem, the interface regularization length scale and the out-of-bounds velocity scale are defined by $\epsilon=\Delta x = 0.01$ and $\Gamma^*=2.5$, respectively.


%
\begin{figure}
    \centering
    \includegraphics[width=0.95\textwidth]{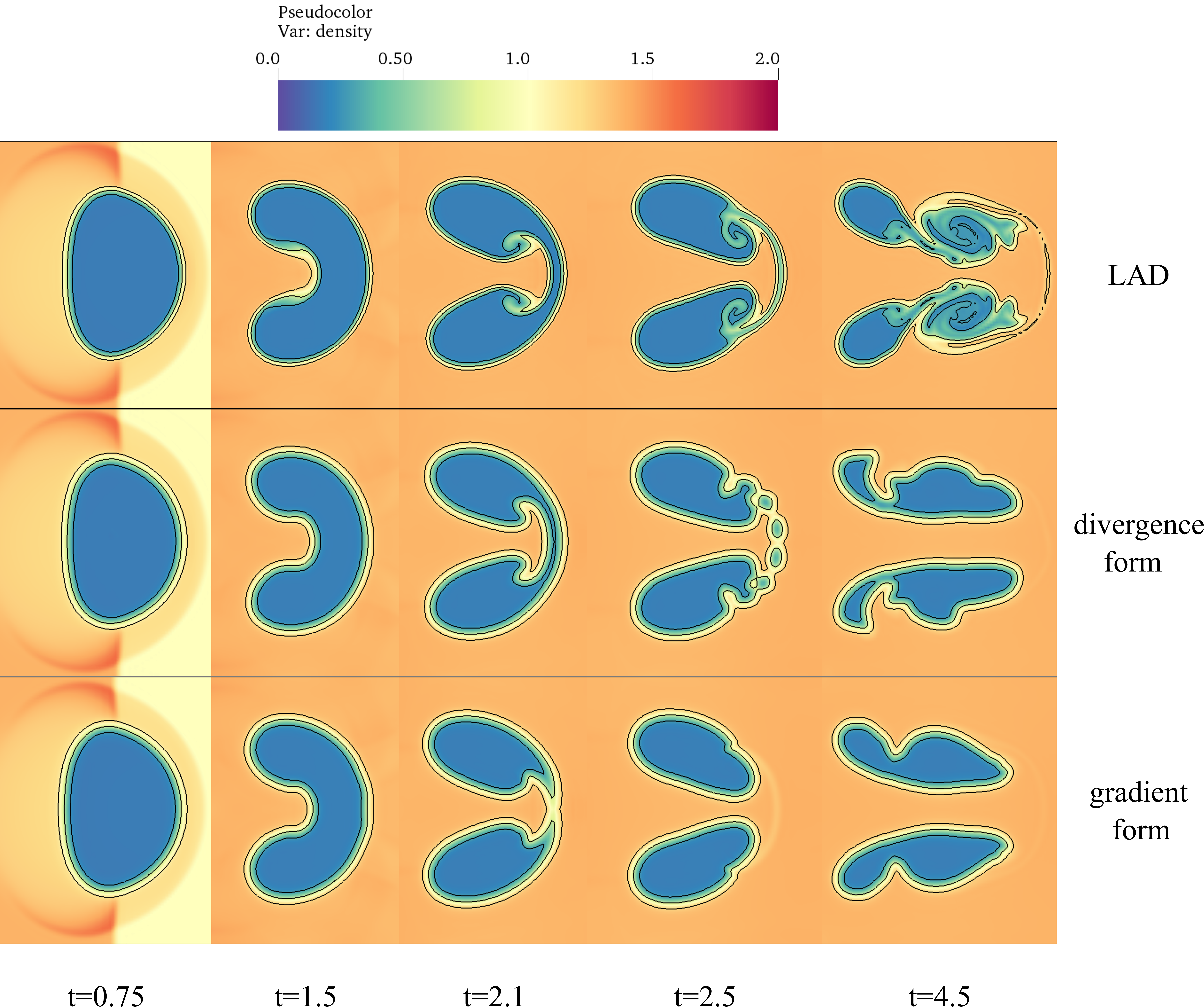}
    \caption{Comparison of the bubble shapes at different times for the case of the shock/helium-bubble-in-air interaction using various interface-capturing methods. The three solid black lines denote the isocontours of the volume fraction values of 0.1, 0.5, and 0.9, representing the interface region.}
    \label{fig:helium_bubble}
\end{figure}
\begin{figure}
         \includegraphics[width=0.48\textwidth]{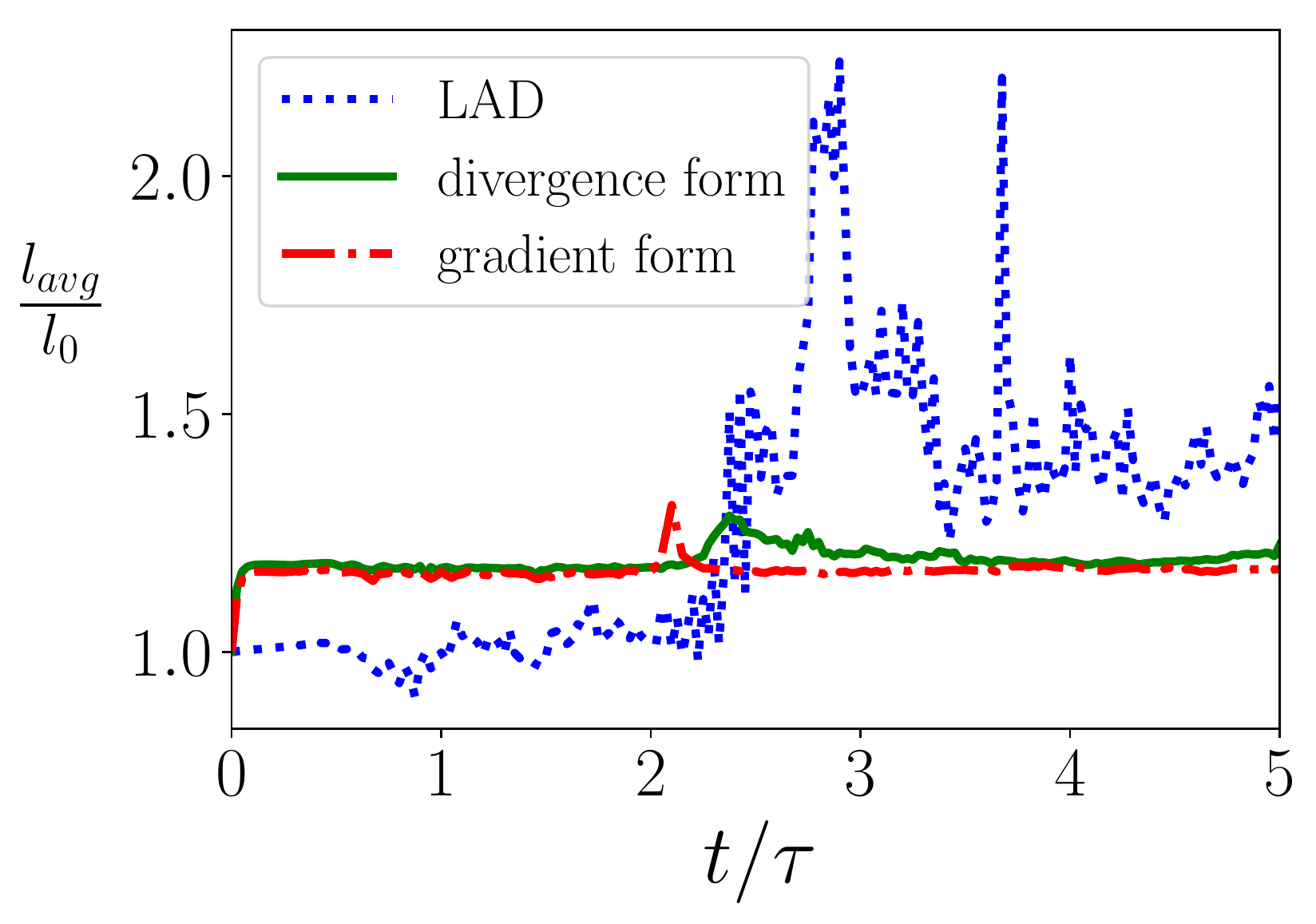}
         \includegraphics[width=0.48\textwidth]{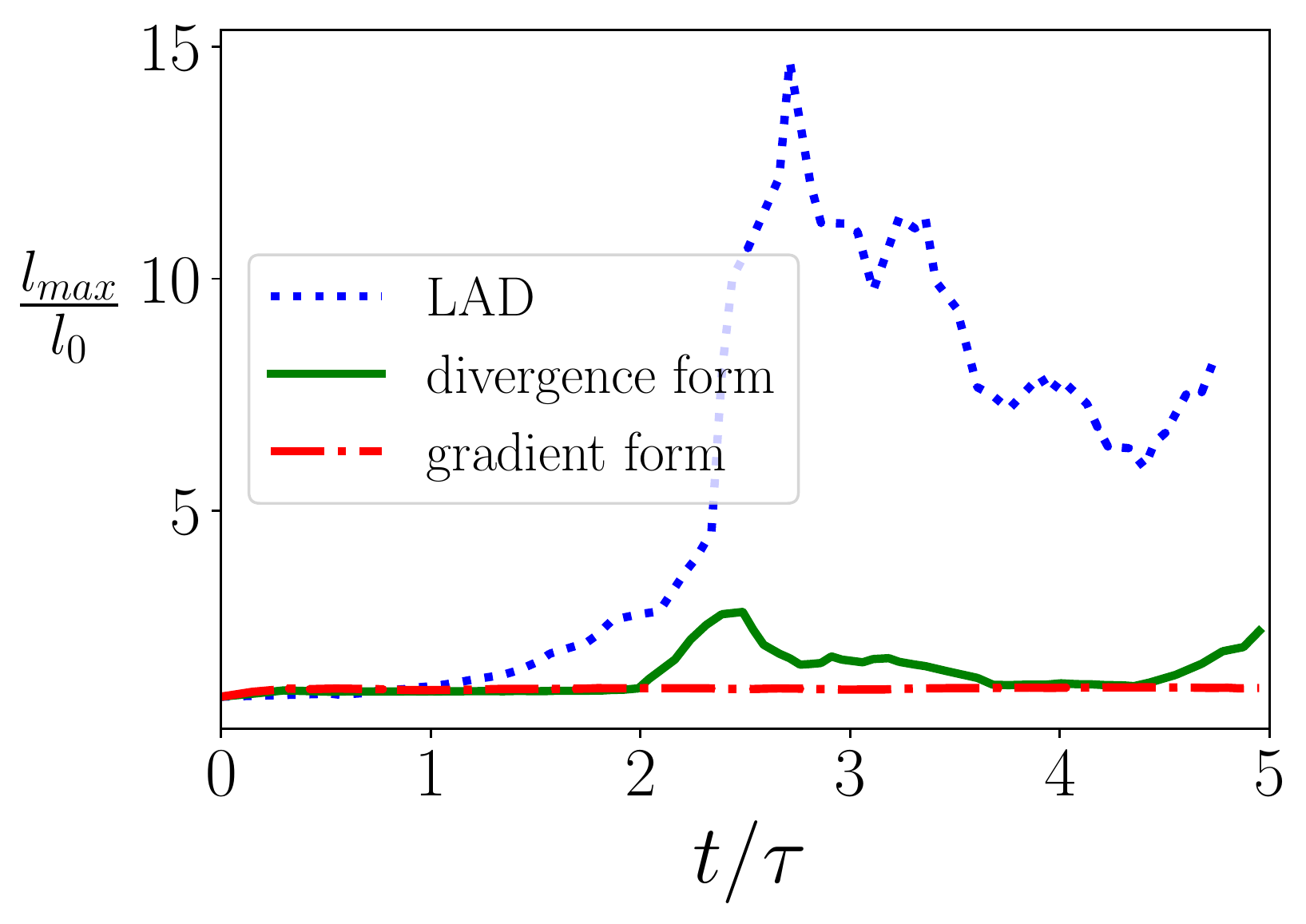}
    \caption{Comparison of the interface-thickness indicator, $l$, by various methods, where $l_0$ is the maximum interface thickness at time $t=0$. (a) Average interface thickness $l_{avg}$. (b) Maximum interface thickness $l_{max}$. To exclude unphysical spikes in $l_{max}$ during breakup events, a moving average of 5 local minima of $l_{max}$ is plotted.}
    \label{fig:helium_thickness}
\end{figure}




The interaction of the shock with the helium bubble and the eventual breakup of the bubble are shown in Figure \ref{fig:helium_bubble}, with depictions of the evolution at various times, for LAD, divergence-form, and gradient-form approaches.
The bubble can be seen to undergo breakup at an approximate (non-dimensional) time of $t=2.5$.
After this time, the simulation cannot be considered physical because of the under-resolved processes associated with the breakup and the lack of explicit subgrid models for these processes;
each interface regularization approach treats the under-resolved processes differently.
Therefore, there is no consensus on the final shape of the bubble among the three methods.
Yet, a qualitative comparison between the three methods can still be made using the results presented in Figure \ref{fig:helium_bubble}. 

Using the LAD approach, the interface diffuses excessively in the regions of high shear, unlike the divergence-form and gradient-form approaches, where the interface thickness is constant throughout the simulation.
However, using the LAD approach, the interface remains sharp in the regions where there is no shearing.
To quantify the amount of interface diffusion, the interface-thickness indicator [$l$ of Eq.~\eqref{equ:thick}] is plotted in Figure \ref{fig:helium_thickness} for the three methods. The average thickness, $l_{avg}$, increases slightly for the LAD method around $t/\tau\approx 2$, but the maximum interface thickness, $l_{max}$, increases almost $15$ times for the LAD method, whereas it remains on the order of one for the other two methods.
This demonstrates a deficiency of the LAD approach for problems that involve significant shearing at an interface that is not subjected to compression. 


Furthermore, the behavior of bubble breakup is significantly different among the various methods.
Depending on the application, any one of these methods may or may not result in an appropriate representation of the under-resolved processes.
However, for the current study that involves modeling interfaces between immiscible fluids, the grid-induced breakup of the divergence-form approach may be more suitable than the diffusion of the fine structures in the LAD approach or the premature loss of fine structures and associated conservation error of the gradient-form approach.
For the LAD approach, the thin film formed at around time $t=2.1$ does not break; rather, it evolves into a thin region of well-mixed fluid.
This behavior may be considered unphysical for two immiscible fluids, for which the physical interface is infinitely sharp in a continuum sense;
this behavior would be more appropriate for miscible fluids.
For the divergence-form approach, the thin film forms satellite bubbles, which is expected when there is a breakage of a thin ligament between droplets or bubbles due to surface-tension effects.
However, this breakup may not be considered completely physical without any surface-tension forces, because the breakup is triggered by the lack of grid support.
For the gradient-form approach, the thin film formed at around time $t=2.1$ breaks prematurely and disappears with no formation of satellite bubbles, and the mass of the film is lost to conservation error.

In Figure 2 of \citet{shukla2010interface}, without the use of interface regularization terms, the interface thickness is seen to increase significantly.
Their approach without interface regularization terms is most similar to our LAD approach, because the LAD approach does not include any sharpening terms.
Therefore, comparing these results suggests that the thickening of the interface in their case was due to the use of the more dissipative Riemann-solver/reconstruction scheme.
The results from the gradient-form approach also match well with the results of the similar method shown in Figure 2 of \citet{shukla2010interface}, which further verifies our implementation.
Finally, there is no consensus on the final shape of the bubble among the three methods, which is to be expected, because 
there are no surface-tension forces and the breakup is triggered by the lack of grid resolution.

\begin{figure}
    \centering
         \includegraphics[width=0.6\textwidth]{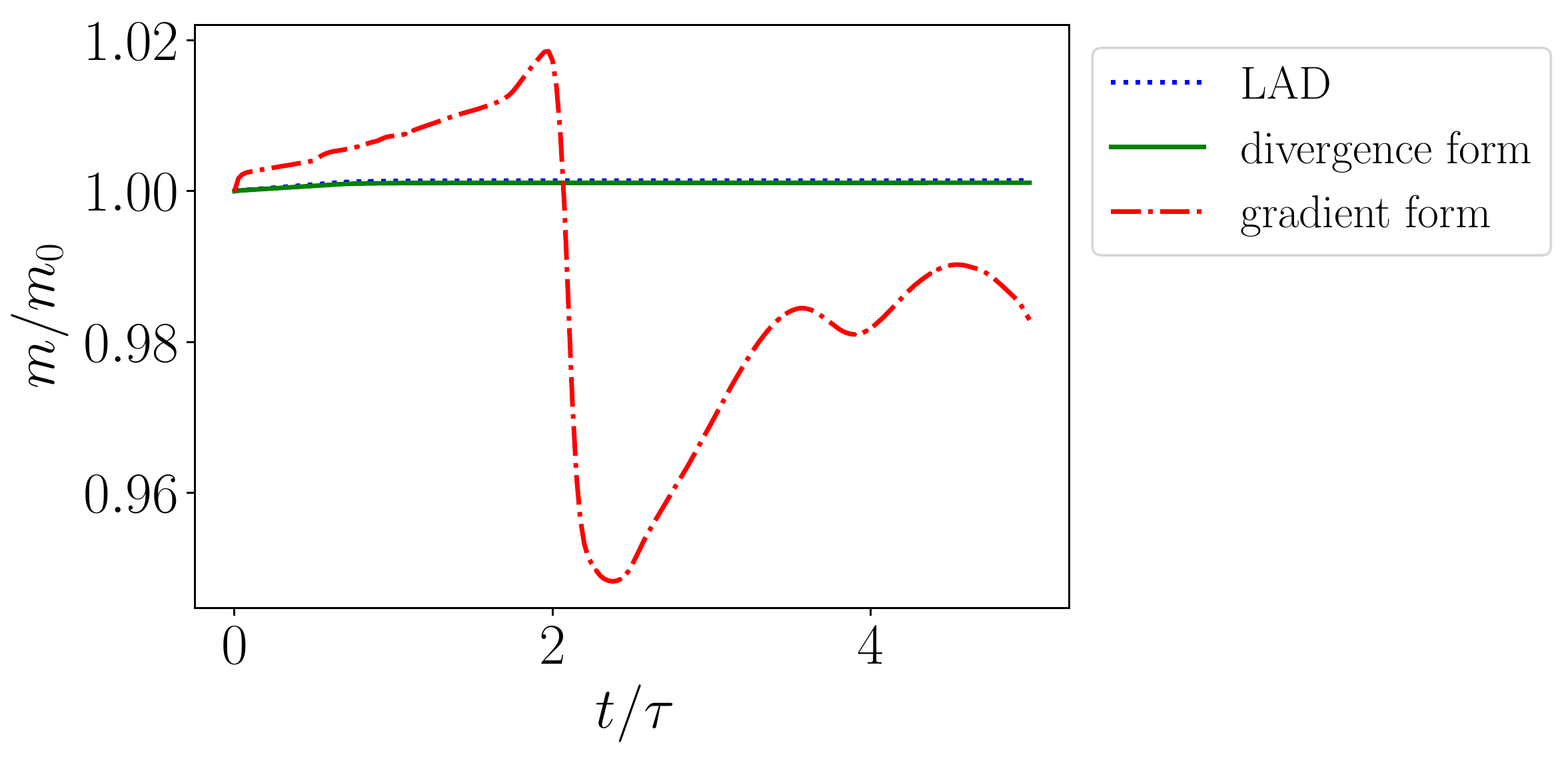}
    \caption{Plot of total mass, $m$, of the helium bubble by various methods, where $m_0$ is the mass at time $t=0$.}
    \label{fig:helium_mass}
\end{figure}

To further quantify the amount of mass lost or gained, the total mass of the bubble is computed using Eq.~\eqref{equ:mass} and is plotted over time in Figure \ref{fig:helium_mass}.
The mass of the bubble is conserved for the LAD and divergence-form approaches, but is not conserved for the gradient-form approach, as expected.
The loss of mass is observed to be largest when the bubble is about to break, for the gradient form approach.
This is because the mass-conservation error in the gradient-form approach is proportional to the local curvature, as described in Section 2.7. Therefore, at the onset of breakup, thin film rupture is different from the other two methods, and the satellite bubbles are absent.

\subsection{Shock interaction with an air bubble in water \label{sec:bubble_collapse}}



This section examines a shock wave traveling through water followed by an interaction with a stationary air bubble.
The material properties are the same as those described in Section~\ref{sec:bubble_advection}. This test case is based on the experiments in \citet{bourne1992shock} and has been widely used as a validation case for various numerical methods and models such as a front-tracking method in \citet{terashima2009front}; a level-set method in \citet{hu2004interface,nourgaliev2006adaptive}; a ghost-fluid method in \citet{bai2017sharp}; a volume-of-fluid method in \citet{bo2014volume}; an implicit diffuse-interface method with a Godunov scheme in \citet{ansari2013numerical}, and with a TENO scheme in \citet{haimovich2017numerical}; and with a gradient-form diffuse-interface approach in \citet{shukla2010interface,shukla2014nonlinear}.

The initial conditions for the velocity, pressure, volume fraction, and density are
\begin{equation}
    \begin{gathered}
        u=u^{(2)} f_s + u^{(1)} \left(1-f_s\right), \quad
        v=0, \quad
        p=p^{(2)} f_s + p^{(1)} \left(1-f_s\right), \\
        \phi_1 = \phi_1^{\epsilon} + \left(1-2\phi_1^{\epsilon}\right) f_{\phi}, \quad \phi_2 = 1-\phi_1, \quad
        \rho = \left(\phi_1 \rho_1 + \phi_2 \rho_2\right) \left[ \rho^{(2)}/\rho^{(1)} f_s  +  \left(1-f_s\right) \right], \\
    \end{gathered}
\end{equation}
respectively, in which the volume fraction function, $f_{\phi}$, and the shock function, $f_s$, are given by,
\begin{equation}
    f_{\phi} =\frac{1}{2} \left\{1 -\textrm{erf} \left[\frac{1-\left(x-2.375\right)^2 - \left(y-2.5\right)^2}{\Delta x}\right] \right\}\quad \textrm{and} \quad f_s = \frac{1}{2} \left[ 1 - \textrm{erf}\left(\frac{x+1}{10 \Delta x}\right) \right],
\end{equation}
respectively, with jump conditions across the shock for velocity $\left(u^{(1)}=0;\,u^{(2)}=68.5176\right)$, density $\left(\rho^{(1)}=1\right.$, $\left.\rho^{(2)}=1.32479\right)$, and pressure $\left(p^{(1)}=1;\,p^{(2)}=19150\right)$.
The problem domain spans $\left(-2\leq x \leq 8;\, 0\leq y\leq 5\right)$, with periodic boundary conditions in the $y$ direction.
A symmetry boundary is applied at $x=8$, representing a perfectly reflecting wall, and a sponge boundary condition is applied over $\left(-2\leq x \leq -1.5\right)$, modeling a non-reflecting free boundary.
The problem is discretized on a uniform Cartesian grid of size $N_x = 400$ and $N_y = 200$.

For this problem, the artificial bulk viscosity, artificial thermal conductivity, artificial diffusivity, interface regularization length scale, interface regularization velocity scale, and out-of-bounds velocity scale are defined by $C_{\beta}=20$, $C_{\kappa}=0.1$, $C_{D}=20$, $\epsilon=\Delta x = 2.5\times10^{-2}$, $\Gamma=2.0$, and $\Gamma^*=0.0$, respectively.
A fourth-order, penta-diagonal, Pad{\'e} filter is employed for dealiasing in this problem to improve the stability of the shock/bubble interaction. The linear system defining this filter is given by 
\begin{equation}
    \hat{f}_i + \alpha (\hat{f}_{i+1} + \hat{f}_{i-1} ) + \frac{1 - 2 \alpha}{14}(\hat{f}_{i+2} + \hat{f}_{i-2} )  = 
    \frac{4 + 6 \alpha}{7} f_i + 
    \frac{2 + 3 \alpha}{7} (f_{i+1}+f_{i-1})
\end{equation}
where $\alpha=0.499$.
 
\begin{figure}
    \centering
    \includegraphics[width=0.9\textwidth]{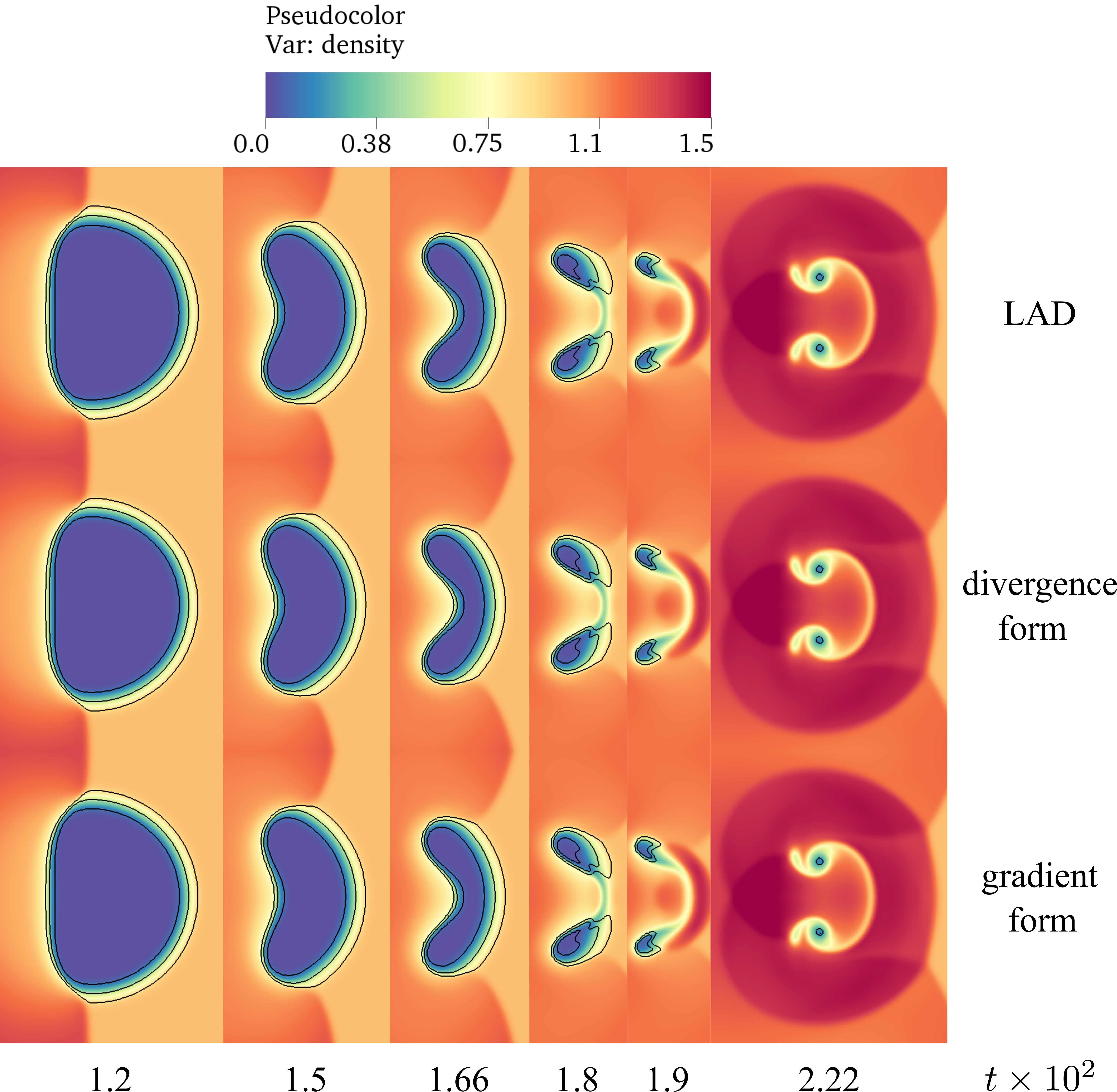}
    \caption{Comparison of the bubble shapes at different times for the case of shock/air-bubble-in-water interaction using various interface-capturing methods. The three solid black lines denote the isocontours of the volume fraction values of 0.1, 0.5, and 0.9, representing the interface region.}
    \label{fig:cavitycollapse}
\end{figure}

Notably, for this problem, the LAD in the mass equations is also necessarily included in the divergence-form and gradient-form approaches to maintain stability.
The latter approaches become unstable for this problem for large $\Gamma$ (the velocity scale for interface regularization).
Figure~\ref{fig:cavitycollapse} describes the evolution in time of the shock/bubble interaction and the subsequent bubble collapse.
There is no significant difference between the various regularization methods for this problem.
The similarity is due to the short convective timescale of the flow relative to the maximum stable timescale of the volume fraction regularization methods; effectively, all methods remain qualitatively similar to the LAD approach.
%

\subsection{Richtmyer--Meshkov instability of a copper--aluminum interface \label{sec:RMI}} 
%
%

\begin{figure}
    \centering
    \includegraphics[width=0.95\textwidth]{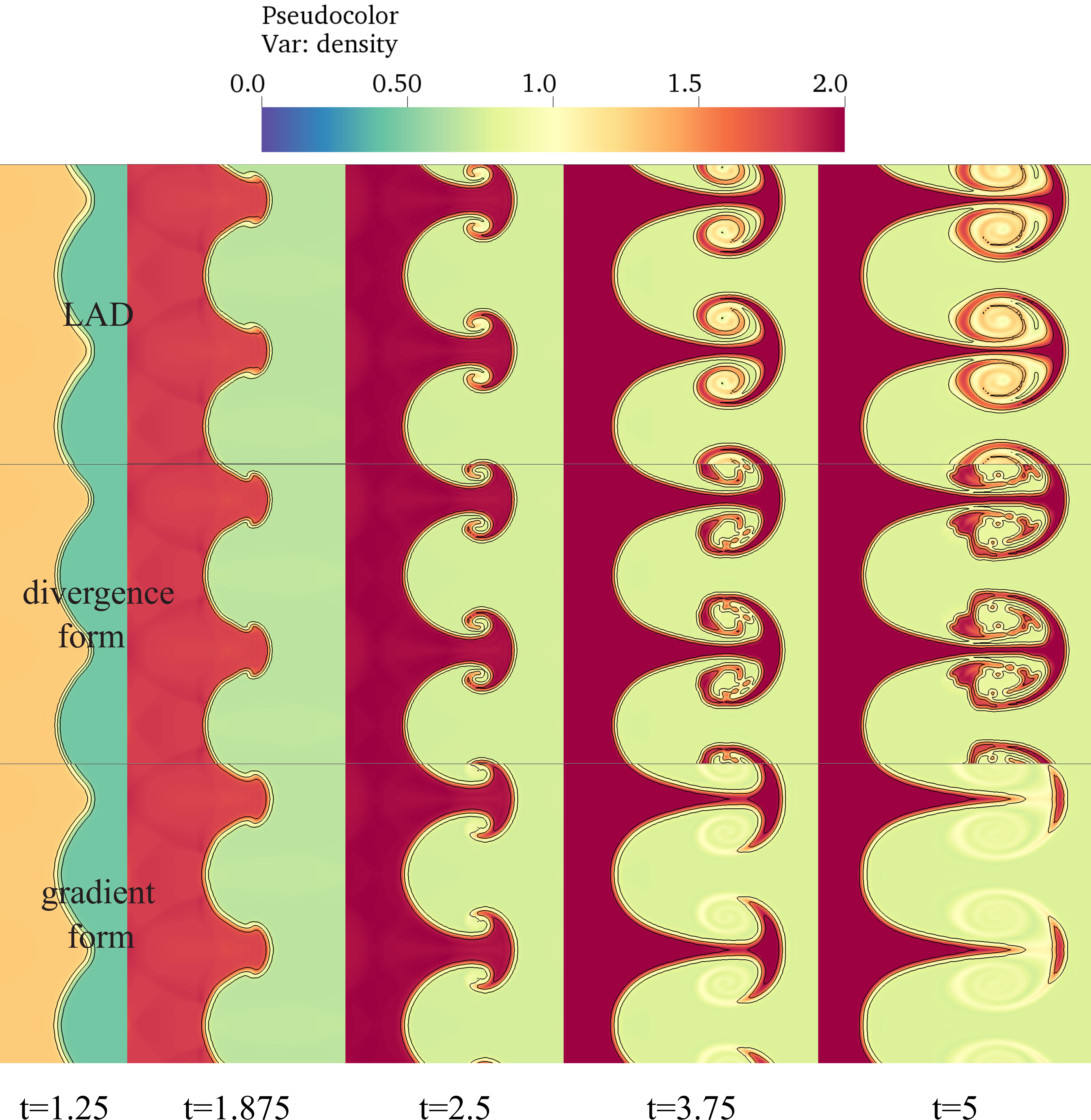}
    \caption{Comparison of the copper--aluminum interface shapes at different times for the Cu-Al RMI case using various interface-capturing methods. The three solid black lines denote the isocontours of the volume fraction values of 0.1, 0.5, and 0.9, representing the interface region.}
    \label{fig:Cu_Al_rmi}
\end{figure}
\begin{figure}
\centering
     \includegraphics[width=0.48\textwidth]{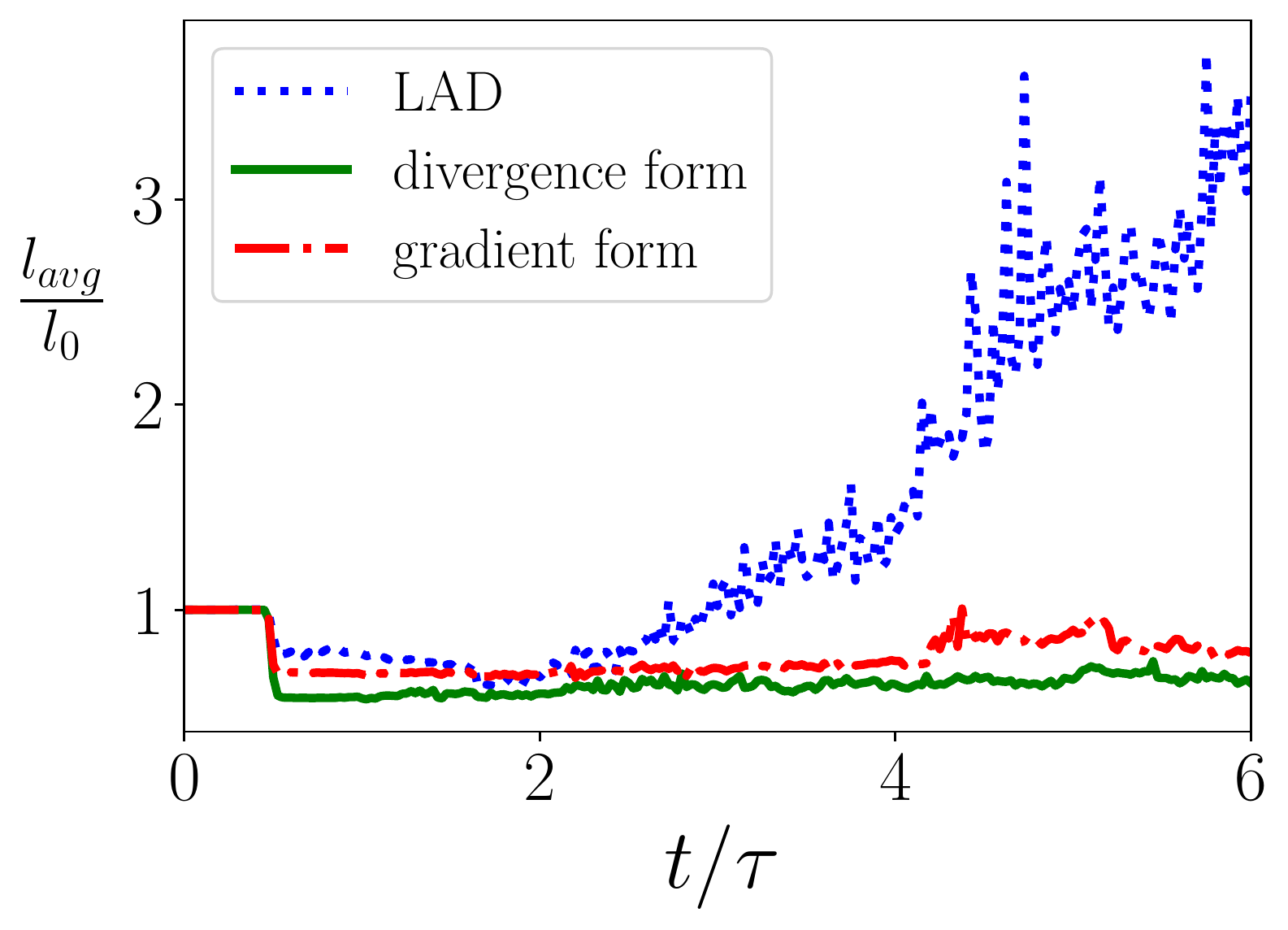}
     \includegraphics[width=0.48\textwidth]{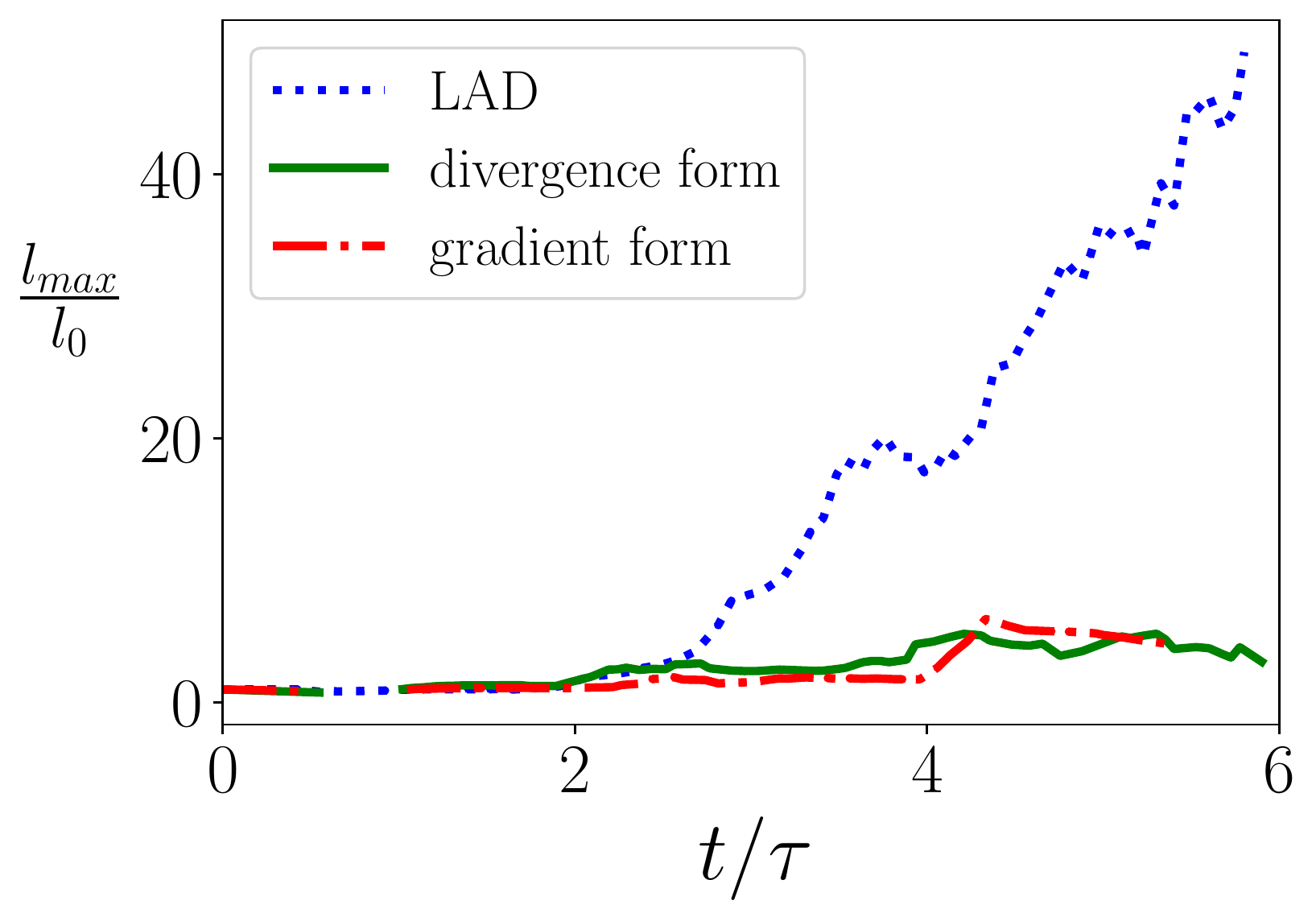}
    \caption{Comparison of the interface-thickness indicator, $l$, by various methods, where $l_0$ is the maximum interface thickness at time $t=0$. (a) Average interface thickness $l_{avg}$. (b) Maximum interface thickness $l_{max}$. To exclude unphysical spikes in $l_{max}$ during breakup events, a moving average of 5 local minima of $l_{max}$ is plotted.}
    \label{fig:rmi_interfaceThickness}
\end{figure}






This section examines a shock wave traveling through copper followed by an interaction with a sinusoidally distorted copper--aluminum material interface. Though this problem has not been as widely studied as the previous examples, it is included to demonstrate how interface regularization methods perform when extended to problems involving elastic-plastic deformation at material interfaces. Such deformation may arise in impact welding, where interfacial instabilities are known to develop as metal plates impact and shear \citep{Nassiri_2016JMPS_Shear}; as well as material characterization at high strain rates, which typically employ a metal-gas configuration of the Richtmyer-Meshkov instability \citep{Dimonte_2011PRL_Use}. The copper-aluminum variant of this problem was previously studied by \cite{lopez2013simulation}, who used a level-set method combined with the modified ghost-fluid method to set boundary conditions at material interfaces. This problem was also studied by \cite{subramaniam2018high} and \cite{adler2019strain}, and the results presented here are an extension of that work.

The problem domain spans $\left(-2\leq x \leq 4;\, 0\leq y\leq 1\right)$, with periodic boundary conditions in the $y$ direction.
A symmetry boundary is applied at $x=4$, representing a perfectly reflecting wall, and a sponge boundary condition is applied over $\left(-2\leq x \leq -1.5\right)$, modeling a non-reflecting free boundary.
The problem is discretized on a uniform Cartesian grid of size $N_x = 768$ and $N_y = 128$.
The material properties for the copper medium are described by $\gamma_1=2.0$, $\rho_1= 1.0$, ${p_{\infty}}_1=1.0$, $\mu_1 = 0.2886$, and ${\sigma_Y}_1 = 8.79\times10^{-4}$.
The material properties for the aluminum medium are described by $\gamma_2=2.088$, $\rho_2= 0.3037$, ${p_{\infty}}_2 = 0.5047$, $\mu_2 = 0.1985$, and ${\sigma_Y}_2 = 2.176\times10^{-3}$.

The initial conditions for the velocity, pressure, volume fraction, and density are
\begin{equation}
    \begin{gathered}
        u=u^{(2)} f_s + u^{(1)} \left(1-f_s\right), \quad
        v=0, \quad
        p=p^{(2)} f_s + p^{(1)} \left(1-f_s\right),\\
        \phi_1 = \phi_1^{\epsilon} + \left(1-2\phi_1^{\epsilon}\right) f_{\phi}, \quad
        \phi_2 = 1-\phi_1, \quad
        \rho = \left(\phi_1 \rho_1 + \phi_2 \rho_2\right) \left[ \rho^{(2)}/\rho^{(1)} f_s  +  \left(1-f_s\right) \right],
    \end{gathered}
\end{equation}
respectively, in which the volume fraction function, $f_{\phi}$, and the shock function, $f_s$, are given by
\begin{equation}
    f_{\phi} =\frac{1}{2} \left(1 -\textrm{erf} \left\{\frac{x- \left[2+0.4/\left(4\pi y\right)\sin\left(4\pi y\right) \right]}{3\Delta x}\right\} \right)\quad \textrm{and} \quad f_s = \frac{1}{2} \left[ 1 - \textrm{erf}\left(\frac{x-1}{2 \Delta x}\right) \right],
\end{equation}
respectively, with jump conditions across the shock for velocity $\left(u^{(1)}=0;\,u^{(2)}=0.68068\right)$, density $\left(\rho^{(1)}=1\right.$, $\left.\rho^{(2)}=1.4365\right)$, and pressure $\left(p^{(1)}=5\times10^{-2};\,p^{(2)}=1.25\right)$.
The kinematic tensors are initialized in a pre-strained state consistent with the material compression associated with shock initialization, assuming no plastic deformation has yet occurred, with
\begin{equation}
    g_{ij} = g^e_{ij} = \begin{cases}
        \left[\rho^{(2)} f_s + \rho^{(1)} \left(1-f_s\right) \right]/\rho_1, & \text{for } i=j=1 \\
        \delta_{ij}, & \text{else }
      \end{cases} \qquad \textrm{and} \qquad g^p_{ij} = \delta_{ij}.
\end{equation}
For this problem, the interface regularization length scale and the out-of-bounds velocity scale for the divergence form method are defined by $\epsilon=\Delta x /2 = 3.90625 \times10^{-3}$ and $\Gamma^*=1.0$, respectively. For the gradient form method, it is necessary for stability to use $\epsilon=3\Delta x/4 = 5.859375 \times10^{-3}$, $\Gamma^*=1.0$, and $\phi_{min} = 1 \times10^{-5}$.





The time evolution of the growth of the interface instability is shown in Figure \ref{fig:Cu_Al_rmi}.
The simulation is integrated well into the nonlinear regime where the bubble (lighter medium) and the spike (heavier medium) have interpenetrated, forming mushroom-shaped structures with fine ligaments.
The qualitative comparison between the methods in this test case is similar to that of the shock-helium-bubble interaction in air.
With the LAD approach, the interface thickness increases with time, especially in the regions of high shear at the later stages.
However, with the divergence-form and gradient-form approaches, the interface thickness is constant throughout the simulation.
This is quantified by plotting the interface-thickness indicator [$l$ of Eq. \eqref{equ:thick}] for each of the three methods in Figure \ref{fig:rmi_interfaceThickness}. The average thickness, shown in Figure \ref{fig:rmi_interfaceThickness}(a) shows a sharp drop in thickness at $t/\tau \approx 0.5$ when the shock passes through the interface. After this, the thickness remains small for both the gradient and divergence form methods, whereas with LAD the interface thickness grows gradually after $t/\tau \approx 2$, when the interface begins to roll up. Figure \ref{fig:rmi_interfaceThickness}(b) shows the maximum interface thickness, which increases almost $60$ times for the LAD method, whereas it stays on the order of one for the other two methods. This illustrates that the LAD method incurs significant artificial diffusion when the interface deformation cannot be resolved by the grid.

\begin{figure}
     \centering
     \includegraphics[width=0.48\textwidth]{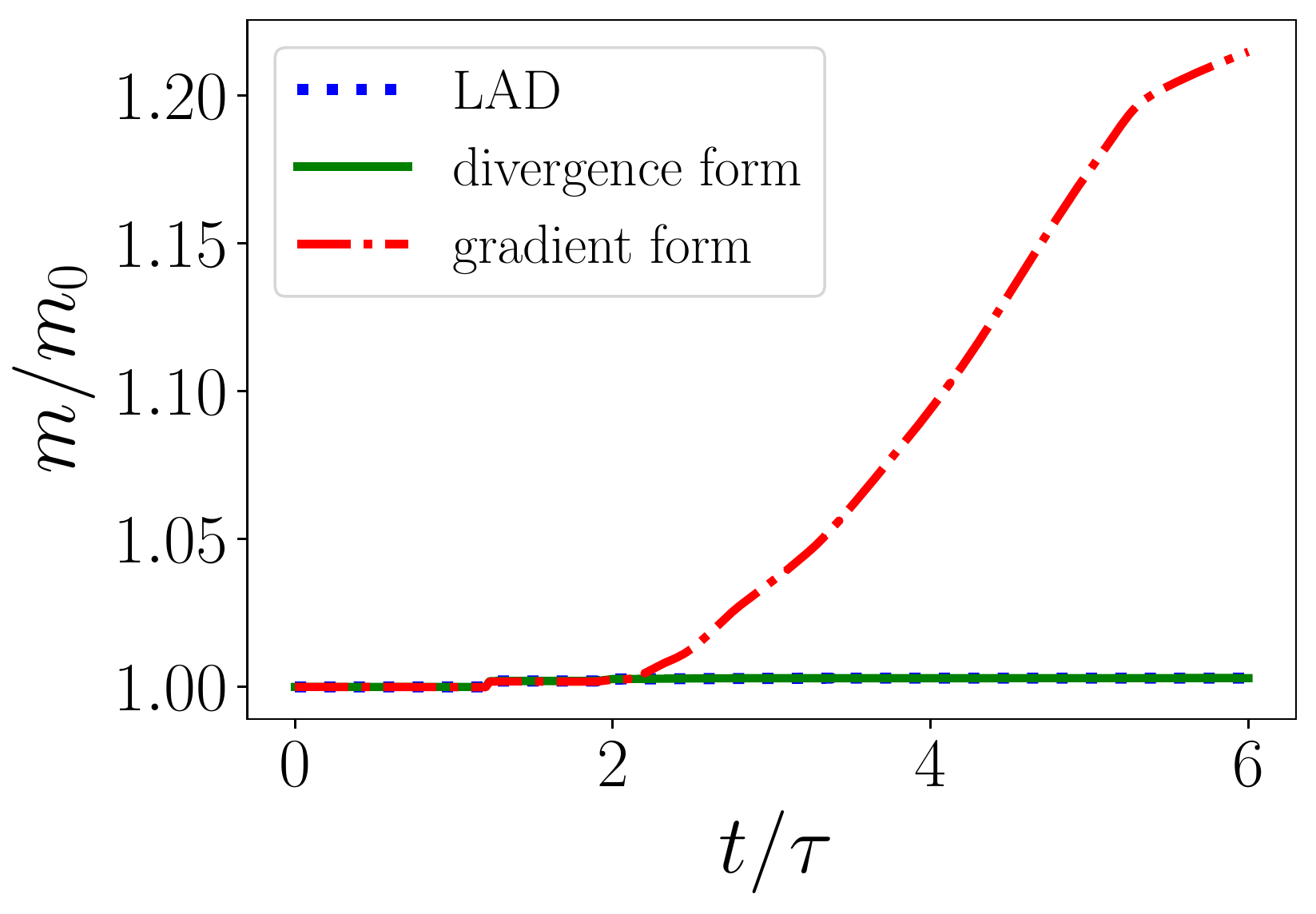}
    \caption{Plot of total mass, $m$, of aluminum by various methods, where $m_0$ is the mass at time $t=0$.}
    \label{fig:rmi_Almass}
\end{figure}

It is also evident from Figure \ref{fig:Cu_Al_rmi} that the gradient-form approach results in significant copper mass loss, and the dominant mushroom structure formed in the nonlinear regime is completely lost.
To quantify the amount of mass lost or gained, the total mass of the aluminum material [Eq. \eqref{equ:mass}] is plotted against time in Figure \ref{fig:rmi_Almass}.
The gradient-form approach results in significant gain in the mass of the aluminum material, up to $20\%$, as the grid can no longer resolve the increased interface curvature during roll-up.
This makes it practically unsuitable for accurate interface representation for long-time numerical simulations. 
With the divergence-form approach, the breakage of the ligaments to form metallic droplets can be seen in Figure \ref{fig:Cu_Al_rmi}.

\section{Summary and concluding remarks}\label{sec:conclusions} 
This work examines three diffuse-interface-capturing methods and evaluates their performance for the simulation of immiscible compressible multiphase fluid flows and elastic-plastic deformation in solids.
The first approach is the \textit{localized-artificial-diffusivity} method of \citet{cook2007artificial}, \citet{subramaniam2018high}, and \citet{adler2019strain}, in which artificial diffusion terms are added to the individual phase mass fraction transport equations and are coupled with the other conservation equations.
The second approach is the \textit{gradient-form approach} that is based on the quasi-conservative method of \citet{shukla2010interface}.
In this method, the diffusion and sharpening terms (together called regularization terms) are added to the individual phase volume fraction transport equations and are coupled with the other conservation equations \citep{tiwari2013diffuse}.
The third approach is the \textit{divergence-form approach} that is based on the fully conservative method of \citet{jain2020conservative}.
In this method, the diffusion and sharpening terms are added to the individual phase volume fraction transport equations and are coupled with the other conservation equations. 
In the present study, all of these interface regularization methods are used in conjunction with a four-equation, multicomponent mixture model, in which pressure and temperature equilibria are assumed among the various phases.
The latter two interface regularization methods are commonly used in the context of a five-equation model, in which temperature equilibrium is not assumed.

The primary objective of this work is to compare these three methods in terms of their ability to: maintain constant interface thickness throughout the simulation; conserve mass of each of the phases, mixture momentum, and total energy; and maintain accurate interface shape for long-time integration.
The secondary objective of this work is to extend these methods for modeling the interface between deforming solid materials with strength.
The LAD method has previously been used for simulating material interfaces between solids with strength \citep{subramaniam2018high, adler2019strain}.
Here, we introduce consistent corrections in the kinematic equations for the divergence-form and the gradient-form approaches to extend these methods for the simulation of interfaces between solids with strength. 

\begin{table}
\centering
\begin{adjustbox}{max width=\textwidth}
\begin{tabular}{@{}|c|c|c|c|c|@{}}
\toprule
Method & \begin{tabular}[c]{@{}c@{}}Conservation \\ \end{tabular} &
\begin{tabular}[c]{@{}c@{}}Sharp \\ interface\end{tabular} &
\begin{tabular}[c]{@{}l@{}}Shape \\ preservation\end{tabular} & \begin{tabular}[c]{@{}l@{}} Behavior of under-resolved\\ ligaments and breakup\end{tabular} \\ \midrule
\begin{tabular}[c]{@{}c@{}}LAD\end{tabular} & Yes & \begin{tabular}[c]{@{}c@{}}{\underline{No}}\\ (interface\\ diffuses\\ in the regions\\ of high shear)\end{tabular} & Yes & \begin{tabular}[c]{@{}c@{}}{\underline{Artificial diffusion}}\\ (fine-scale features artificially\\ diffuse as they \\ approach unresolved scales)\end{tabular} \\ \midrule
\begin{tabular}[c]{@{}c@{}}Divergence\\ form\end{tabular} & Yes & Yes & \begin{tabular}[c]{@{}c@{}}{\underline{No}}\\ (interface\\ aligns with\\ the grid)\end{tabular} & \begin{tabular}[c]{@{}c@{}}{\underline{Artificial breakup}}\\ (fine-scale features artificially \\ break up as they\\ approach unresolved scales)\end{tabular} \\ \midrule
\begin{tabular}[c]{@{}c@{}}Gradient\\ form\end{tabular} & \begin{tabular}[c]{@{}c@{}}{\underline{No}}\\ (under-resolved\\ features\\ will be lost)\end{tabular} & Yes & Yes & \begin{tabular}[c]{@{}c@{}}{\underline{Artificial loss of mass}}\\ (fine-scale features are lost, \\ due to conservation error, as they \\ approach unresolved scales)\end{tabular} \\ \bottomrule
\end{tabular}
\end{adjustbox}
\caption{Summary of the advantages and disadvantages of the three diffuse-interface capturing methods considered in this study: LAD method based on \citet{cook2007artificial}, \citet{subramaniam2018high}, and \citet{adler2019strain}; divergence-form approach based on \citet{jain2020conservative}; and the gradient-form approach based on \citet{shukla2010interface} and \citet{tiwari2013diffuse}. The relative disadvantages of each approach and the different behaviors of under-resolved processes are underlined.}
\label{tab:summary}
\end{table}

We employ several test cases to evaluate the performance of the methods, including (1) advection of an air bubble in water, (2) shock interaction with a helium bubble in air, (3) shock interaction and the collapse of an air bubble in water, and (4) Richtmyer--Meshkov instability of a copper--aluminum interface.
For the application of these methods to large-scale simulations of engineering interest, it is rarely practical to use hundreds of grid points to resolve the diameter of a bubble/drop.
Therefore, we choose to study the limit of relatively coarse grid resolution, which is more representative of the true performance of these methods.

The performance of the three methods is summarized in Table~\ref{tab:summary}.
The LAD and the divergence-form approaches conserve mass, momentum, and energy, whereas the gradient-form approach does not.
The mass-conservation error increases proportionately with the local interface curvature; therefore, fine interfacial structures will be lost during the simulation.
The divergence-form and the gradient-form approaches maintain a constant interface thickness throughout the simulation, whereas the interface thickness of the LAD method increases in the regions of high shear due to the lack of interface sharpening terms to counter the artificial diffusion.
The LAD and the gradient-form approaches maintain the interface shape for a long time compared to the divergence-form approach; however, the interface distortion of the divergence-form approach can be mitigated with the use of appropriately crafted higher-order schemes for the interface regularization terms.

For each method, the behavior of under-resolved ligaments and breakup features is unique. 
For the LAD approach, thin ligaments that form at the onset of bubble breakup (or in late-stage RMI) diffuse instead of rupturing.
For the gradient-form approach, the ligament formation is not captured because of mass-conservation issues, which result in premature loss of these fine-scale features.
For the divergence-form approach, the ligaments rupture due to the lack of grid support, acting like an artificial surface tension force that becomes significant at the grid scale.

For broader applications, the optimal method depends on the objectives of the study.
These applications include (1) well-resolved problems, in which differences in the behavior of under-resolved features is not of concern, (2) applications involving interfaces between miscible phases, and (3) applications involving more complex physics, including regimes in which surface tension or molecular diffusion must be explicitly modeled and problems in which phase changes occur.
We intend this demonstration of the advantages, disadvantages, and behavior of under-resolved phenomena exhibited by the various methods to be helpful, albeit being unphysical, in the selection of an interface-regularization method.
These results also provide motivation for the development of subgrid models for multiphase flows.

\section*{Acknowledgments} 
S.~S.~J. was supported by a Franklin P. and Caroline M. Johnson Stanford Graduate Fellowship. M.~C.~A. and J.~R.~W. appreciate the sponsorship of the U.S. Department of Energy Lawrence Livermore National Laboratory under contract DE-AC52-07NA27344 (monitor: Dr. A.~W. Cook). Authors also acknowledge the Predictive Science Academic Alliance Program III at Stanford University. A preliminary version of this work has been published \citep{adler2020diffuse,jain2020diffuse} as the Center for Turbulence Research Annual Research Briefs (CTR-ARB) and are available online\footnote{http://web.stanford.edu/group/ctr/ResBriefs/2020/32\_Adler.pdf}\footnote{http://web.stanford.edu/group/ctr/ResBriefs/2020/33\_Jain.pdf}. S.~S.~J., M.~C.~A., and J.~R.~W. are thankful for Dr. Kazuki Maeda, for reviewing the CTR-ARBs and for his useful comments which helped improve the final version of the article.

\section*{Appendix A: Finite-difference operators for the divergence-form approach}

The test case of shock interaction with a helium bubble in air is repeated for the divergence-form approach with the same parameters listed in Section \ref{sec:helium_bubble}. Here, the difference is in the numerical representation of the nonlinear interface-regularization terms. In Section \ref{sec:helium_bubble}, the interface-regularization fluxes are formed at the faces, as described in Section \ref{sec:numerics}, which is consistent with the finite-volume implementation in \citet{jain2020conservative}. Whereas, here, a second-order standard central finite-difference scheme is used instead.

\begin{figure}
    \centering
    \includegraphics[width=\textwidth]{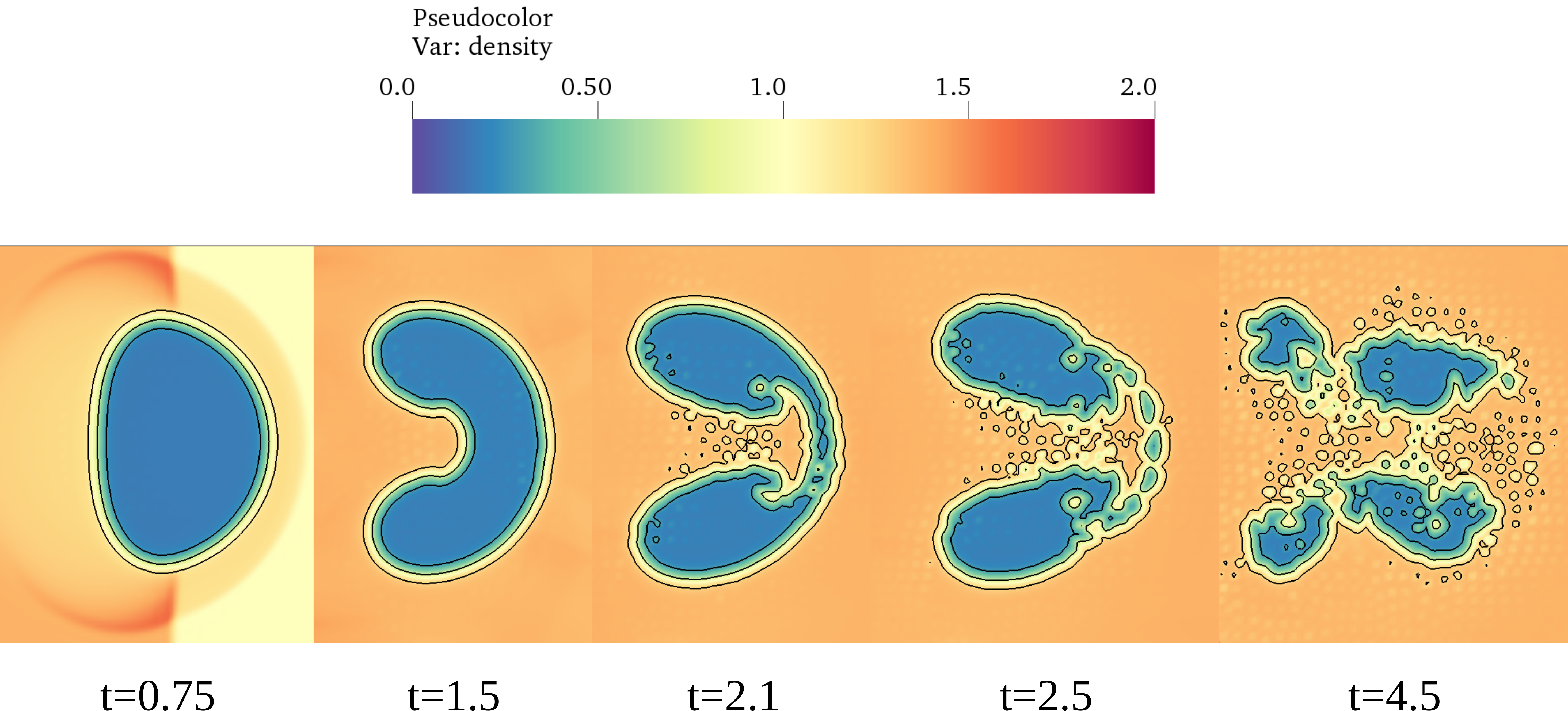}
    \caption{The bubble shapes at different times for the case of the shock/helium-bubble-in-air interaction using the divergence-form approach, where the interface-regularization terms are discretized using a second-order standard central finite-difference scheme. The three solid black lines denote the isocontours of the volume fraction values of 0.1, 0.5, and 0.9, representing the interface region.}
    \label{fig:helium_bubble_FD}
\end{figure}

The shock interaction with the helium bubble in air and the subsequent evolution of the bubble shape are shown in Figure \ref{fig:helium_bubble_FD}. An unphysical wrinkling of the interface can be seen at the later stages of the bubble deformation. This behavior is consistent with the observations made by \citet{shukla2010interface}, which motivated them to develop the gradient-form approach. However, discretizing the fluxes at the faces, \citet{jain2020conservative} showed that this results in discrete balance between the diffusion and sharpening fluxes, thereby eliminating the spurious wrinkling of the interface as can be seen in Figure \ref{fig:helium_bubble}. In this work, this face-evaluated flux formulation has been succesfully extended for higher-order schemes and is presented in Section \ref{sec:numerics}.


%
%


\bibliographystyle{model1-num-names}
\bibliography{papers.bib}

\end{document}